\documentclass[12pt,aps,amsmath,latexsym,amsfonts]{JHEP3}
\usepackage{epsfig}
\usepackage{amsfonts,amssymb,amsmath}
\usepackage[hang]{subfigure}
\usepackage[numbers, square, sort&compress]{natbib}

\newcommand{\be}{\begin{equation}}
\newcommand{\ee}{\end{equation}}
\newcommand{\bea}{\begin{eqnarray}}
\newcommand{\eea}{\end{eqnarray}}

\title{D7-brane dynamics and thermalization\\ in the Kuperstein-Sonnenschein model}
\author{
Dariush Kaviani\thanks{Email:dariush@ipm.ir} \\
School of Particles and Accelerators,
Institute for Research in Fundamental Sciences (IPM),
P.O. Box 19395-5531, Tehran, Iran
}

\abstract{We study the temperature of rotating probe D7-branes, dual
to the temperature of flavored quarks, in the Kuperstein--Sonnenschein
holographic model including the effects of spontaneous breakdown of
the conformal and  chiral flavor symmetry. The model embeds probe 
D7-branes into the Klebanov-Witten gravity dual of conformal gauge
theory, with the embedding parameter, given by the minimal radial
extension of the probe, setting the IR scale of conformal and chiral
flavor symmetry breakdown. We show that when the minimal extension is
positive definite and additional spin is turned on, the induced world
volume  metrics on the probe admit thermal horizons and Hawking 
temperatures despite the absence of black holes in  the bulk. We find
the scale and behavior of the temperature in flavored quarks are 
determined notably by the IR scale of symmetry breaking, and by the 
strength and sort of external fields. We also derive the energy--stress
tensor of the rotating probe and study its backreaction and energy 
dissipation. We show that at the IR scale the backreaction is 
non-negligible and find the energy can flow from the probe to the bulk,
dual to the energy dissipation from the flavor sector into the gauge theory.}

\keywords{D-branes, Brane Dynamics in Gauge Theories}

\preprint{IPM/PA-418}

\begin{document}

\section{Introduction}
The quest for building realistic models of QCD in string theory
motivated the construction of flavored holographic models including
the spontaneous breakdown of the conformal and chiral flavor symmetry.
In gauge/gravity duality, \cite{Maldacena:1997re}, adding
$N_f$ D7-branes to the near horizon limit of $N_c$ D3-branes includes
open strings extended between the two sorts of branes that transform
in the fundamental representation $U(N_c)\times U(N_f)_L$. Thus in the
probe limit, $N_f\ll N_c$, where the backreaction of the additional
branes is negligible, adding $N_f$ D7-branes to the background
corresponds to adding fundamental quarks in the dual gauge theory 
\cite{Karch:2002sh} (see also \cite{Ammon:2015wua}). The addition of a
further stack of $N_f$ anti D7-branes includes anti-quarks that transform
in the fundamental representation of another $U(N_f)_R$ symmetry, the
gauge symmetry on the new stack of branes. The result is a non-Abelian
$U(N_f)_L\times U(N_f)_R$ gauge symmetry on the gravity side,
corresponding to a global chiral symmetry on the field theory side. This
chiral symmetry gets spontaneously broken, when on the gravity side the
D7-branes and anti D7-branes extend from the UV and join smoothly into
a single curved D7-brane in the IR where only one $U(N_f)_D$ factor
survives (cf. \cite{Ammon:2015wua}). 

The prime example of such scenario is the Kuperstein--Sonnenschein
holographic model \cite{Kuperstein:2008cq}. Motivated by the 
Sakai--Sugimoto model, \cite{Sakai:2004cn} (see also \cite{Ammon:2015wua}),
the model embeds a probe D7-brane into the simplest warped Calabi-Yau throat
background, including the Klebanov-Witten (KW) warped conifold geometry,
\cite{Klebanov:1998hh}, $adS_5\times T^{1,1}$ with $T^{1,1}\cong S^3\times S^2$,
dual to $\mathcal{N}=1$ superconformal gauge field theory. The D7-brane
starts from the UV boundary at infinity, bends at minimal extension in the IR,
and ends up at the boundary. The D7-brane thus forms a  U-shape. As the D7-brane
and anti D7-brane differ only by orientation, the probe describes a supersymmetry
(SUSY) breaking D7-brane/anti D7-brane pair, merged in the bulk at minimal
extension. The SUSY breaking pair also guarantees tadpole cancellation on the
transverse $S^2$ by the annihilation of total D7 charge. When the minimal extension
shrinks to zero at the conifold point, the embedding appears as a disconnected 
D7-brane/anti D7-brane pair. The D7-brane then forms a V-shape. In the V-shape
configuration, the induced world volume metric on the D7-brane is that of 
$adS_5\times S^3$ and the dual gauge theory describes the conformal and chiral
symmetric phase. On contrary, in the U-shape  configuration, the induced world
volume metric on the D7-brane has no $adS$ factor and the conformal and chiral
flavor symmetry get broken\footnote{As discussed in $\S$2, this setup cannot be
realized in the well-used $adS_5\times S^5$ background.}.

The Kuperstein--Sonnenschein model has also been extended.
In refs.\,\cite{Sakai:2003wu,Dymarsky:2009cm} the model has been embedded
into the confining KS background \cite{Klebanov:2000hb}\footnote{Also, in 
refs.\,\citep{Ouyang:2003df} holomorphic embeddings of D7-branes into the
Klebanov-Strassler (KS) solution dual to supersymmetric gauge theory without
flavor chiral symmetry breaking have been considered. The backreaction of
D7-branes in KS $\&$ KW has been studied in  refs.\,\citep{Benini:2006hh}.}.
In ref.\,\cite{Ben-Ami:2013lca}, the model has been extended and classified
by studying different types of probe branes embedded into KW, \cite{Klebanov:1998hh},
and ABJM theory, \cite{Aharony:2008ug}. Moreover, in refs.\,\cite{Ihl:2012bm}
the model has been extended to finite temperature and density by embedding the
probe brane into the $adS$ black hole background and considering gauge theory
at finite chemical potential or baryon number density\footnote{We also note
that such holographic setups with probe branes, \cite{Karch:2001cw,Karch:2000gx}, 
have also been constructed at finite temperature and/or density, to model to
model flavor physics \cite{Kruczenski:2003be,Karch:2002sh} and quantum critical
phenomena \cite{Karch:2007pd}(see also \cite{Ammon:2015wua}), complementary to
other works on charged $adS$ black holes \cite{Herzog:2009xv} 
Such phenomena have also been studied at zero temperature in 
refs.\,\cite{O'Bannon:2008bz,PremKumar:2011ag}.};\,--In the presence of $N_f$
flavors of degenerate mass, the gauge theory admits a global 
$U(N_f)\simeq SU(N_f)\times U(1)$ symmetry, where the $U(1)$ charge counts
the net number of quarks. On the gravity side, this global symmetry corresponds
to the $U(N_f)$ gauge symmetry on the world volume of the $N_f$ D-brane probes.
The  conserved currents related to the $U(N_f)$ symmetry of the gauge theory are
dual to the gauge fields on the D-branes. Therefore, introducing a chemical 
potential or non-vanishing baryon density number in the gauge theory corresponds
to turning on the diagonal $U(1)\subset U(N_f)$ gauge field on the D-branes.

However, earlier studies, \cite{Das:2010yw} (see also \cite{Taghavi:2016nnp}),  
of flavored holographic models have shown that flavored quarks can behave
thermally even when the gauge theory itself is dual to pure $adS$ at zero 
temperature. The prime examples of such non-equilibrium systems have been
constructed in ref.\,\cite{Das:2010yw} and involve time-dependent classical
solutions of probe branes. The model constructed in ref.\,\cite{Das:2010yw}
embeds rotating probe branes into the $adS_5\times S^5$ background solution
dual to $\mathcal{N}=4$ gauge theory at finite $U(1)$ R--charge chemical 
potential, $U(1)_R$;\,--In the presence of $N_f$ flavors the $\mathcal{N}=4$
gauge  theory admits a gauge group $SO(6)_R\simeq SO(4)\times U(1)_R$. The
$U(1)_R$ symmetry rotates left-and right-handed quarks oppositely, as the 
$U(1)$ axial symmetry of QCD, and corresponds to  turning on angular momentum
for the flavor probe brane in the dual $adS_5\times S^5$ supergravity solution.
It has been shown in ref.\,\cite{Das:2010yw} that the induced world volume metrics
on the rotating probe branes admit thermal horizons with characteristic Hawking
temperatures in spite of the absence of black holes in the bulk 
(see also \cite{Russo:2008gb}). By gauge/gravity duality, the temperature of
the probe in supergravity corresponds to the temperature in the flavor sector of
the dual gauge theory. It has thus been concluded in ref.\,\cite{Das:2010yw} that
such flavored holographic models including two different temperatures,--one being
the zero bulk temperature in pure $adS$ while the other the non-zero Hawking 
temperature on the probe brane--, exemplify non-equilibrium steady states. However,
by computing the energy--stress tensor of the system, it has been shown in 
ref.\,\cite{Das:2010yw} that the energy from the probe will eventually dissipate
into the bulk. By duality, it has thus been concluded in ref.\,\cite{Das:2010yw}
that the energy from the flavor sector will eventually dissipate into the gauge theory.

Moreover, recently, \cite{Kaviani:2015rxa}, non-equilibrium steady states have been
studied in more general holographic backgrounds, including warped Calabi-Yau throats, 
\cite{Klebanov:1998hh,Klebanov:2000nc,Klebanov:2000hb}, where some supersymmetry and/or 
conformal invariance are broken. In these holographic conifold solutions, the breakdown
of conformal invariance  modifies the $adS$ structure of the gravity dual, and the 
corresponding $\mathcal{N}=1$ gauge theory has RG cascade in the singular UV, 
\cite{Klebanov:2000nc}, and admits confinement and chiral symmetry breaking in the 
regular IR, \cite{Klebanov:2000hb}. In ref.\,\cite{Kaviani:2015rxa}, embeddings of probe
D1-branes into these holographic QCD-like solutions have been considered. The D1-branes
extend along the time and the holographic radial coordinate, and hence are holographically
dual to magnetic monopoles. In addition, the D1-branes have been allowed to rotate about
spheres of the conifold geometry. The brane equations of motion have been solved 
analytically within the linearized approximation and the induced world volume metrics on
rotating probe D1-branes have been computed. It has been shown ref.\,\cite{Kaviani:2015rxa}
that when the supergravity dual is away from the confining IR limit, \cite{Klebanov:2000hb},
the induced world volume metrics on rotating probe D1-branes in the UV solutions, 
\cite{Klebanov:1998hh,Klebanov:2000nc}, admit distinct thermal horizons and Hawking temperatures
despite the absence of black holes in the bulk Calabi-Yau. In the IR limit, \cite{Klebanov:2000hb},
it has been shown ref.\,\cite{Kaviani:2015rxa} that once the angular velocity, dual to R--charge,
approaches the scale of glueball masses, the world volume horizon hits the bottom of the  throat,
such that the entire D1-brane world volume is inside the horizon, obstructing world volume black
hole formation. In the UV  limit, \cite{Klebanov:2000nc}, however, it has been shown 
ref.\,\cite{Kaviani:2015rxa} that the world volume black hole nucleates, with its world volume
horizon changing dramatically with the scale chiral symmetry breaking. In the conformal UV limit,
\cite{Klebanov:1998hh}, it has been shown in ref.\,\cite{Kaviani:2015rxa} that the induced world
volume metric of the rotating probe D1-brane takes the form of the BTZ black hole metric, modulo
the angular coordinate. The related world volume temperature has been found proportional to the
angular velocity, or R--charge, as expected (as in \cite{Das:2010yw}), increasing faster than the
world volume temperature in the non-conformal case.  In the conformal UV limit, \cite{Klebanov:1998hh},
it has also been shown ref.\,\cite{Kaviani:2015rxa} that by turning on a non-trivial SUGRA background
gauge field, the induced world volume horizon is that of the $adS$-Reissner-Nordstr\"om black hole,
modulo the angular coordinate. The related world volume temperature has been found to have two
distinct branches, one that increases and another that decreases with growing horizon size, describing
`small' and `large' black holes, respectively. It has then been concluded in ref.\,\cite{Kaviani:2015rxa} 
that the $\mathcal{N}=1$ gauge theory which itself is at zero temperature couples to monopoles at
finite temperatures, hence producing non-equilibrium steady states, when the theory is away from 
the confining limit. However, by computing the energy-stress tensor and total angular momentum, it
has been shown in ref.\,\cite{Kaviani:2015rxa} that in the IR of the UV--SUGRA solutions, 
\cite{Klebanov:1998hh,Klebanov:2000nc}, the backreaction of the D1-brane to the background is 
non-negligibly large, even for slow rotations, indicating black hole formation and energy dissipation
in the SUGRA background itself.

The aim of this work is to extend such previous analysis and study non-equilibrium
systems and their energy flow in more general and realistic holographic models
embedding higher dimensional probe branes with spontaneous breakdown of the conformal
and chiral flavor symmetry. The model we consider consists of the Kuperstein--Sonnenschein 
holographic model\footnote{The model is summarized in the second paragraph above, and
discussed in more details in $\S$2.}, allowing the probe brane to have, in addition,
conserved angular motion corresponding to finite R--charge chemical potential. The
motivation is the fact that in such model the $adS$ structure of the induced world
volume metric gets modified by  the embedding parameter, i.e., by the minimal extension
of the brane, which sets the IR scale of conformal and chiral flavor symmetry breaking of
the dual gauge theory. The induced world volume metric on the rotating brane, when given
by the black hole geometry, is then expected to give the Hawking temperature on the probe
dual to the temperature of flavored quarks in the gauge theory. Since the gauge theory 
itself is at zero temperature while its flavor sector is at finite temperature, such
systems  constitute novel examples of non-equilibrium steady states in the gauge theory
of  conformal and chiral flavor symmetry breakdown. However, interactions between different
sectors are expected. The energy-stress tensor of the probe brane is then expected to yield
the energy dissipation from the probe into the system, dual to the energy dissipation from
the flavor sector into the gauge  theory. We are also interested in modifying our analysis
by turning on world volume gauge fields on the brane,--including finite baryon chemical 
potential--, corresponding to turning on external fields in the dual gauge theory. The
motivation is the fact that in the presence of such fields the R--symmetry of the gauge
theory gets broken\footnote{We note that earlier studies in gauge/gravity, \cite{Filev:2007gb},
have shown this result.}  and the corresponding modifications in the induced world volume metric
and energy-stress tensor on the probe are expected to reveal new features of thermalization.

The main results we find are as follows. We first show that when the minimal
extension is positive definite and spin is turned on, the induced world volume 
metrics on rotating probe D7-branes in the Kuperstein--Sonnenschein model admit
thermal horizons and Hawking temperatures despite the absence of black holes in
the bulk. We find the scale and behavior of the temperature on D7-branes are
determined, in particular, strongly by the size of the minimal extension and by
the strength and sort of world volume brane gauge fields. By gauge/gravity duality,
we therefore find the scale and behavior of the temperature in flavored quarks
are determined strongly by the IR scale of conformal and chiral flavor
symmetry breaking, and by the strength and sort of external fields. We note that
by considering the backreaction of such solutions to the holographic KW background,
one naturally expects the D7-brane to form a very small black hole in KW, 
corresponding to a locally thermal gauge field theory in the probe limit. Accordingly,
the rotating D7-brane describes a thermal object in the dual gauge field theory. In
the KW background, the system is dual to  $\mathcal{N}=1$ gauge theory coupled to a
quark. Since the gauge theory itself is at zero temperature while the quark is at 
finite temperature, we find that such systems are in non-equilibrium steady states. 
However, we then show that the energy from the flavor sector will eventually dissipate
into the gauge theory. We first find from the energy--stress tensor that at the minimal
extension, at the IR scale of symmetries breakdown, the energy density blows up and hence
show the backreaction in the IR is non-negligible. We then show from the energy--stress
tensor that when the minimal extension is positive definite and spin is turned on, the
energy flux is non-vanishing and find the energy can flow from the brane into the system,
forming, with the large backreaction, a black hole in the system. By gauge/gravity duality,
we thus find the energy dissipation from the flavor sector into the gauge theory.

The paper is organized as follows. In Sec.\,2, we review the basics of the 
Kuperstein-Sonnenschein holographic model. We first write down the specific
form of the background metric suitable for the probe D7-brane embedding. We
then review the solution of the probe D7-brane equation of motion from the
brane action. In Sec.\,3, we modify the model by turning on conserved angular
motion for the probe D7-brane. We first derive the induced world volume metric
and compute the Hawking temperature on the rotating probe D7-brane. We then 
derive the energy--stress tensor on the rotating brane and compute its 
backreaction and energy dissipation. In Secs.\,4--5, we modify our analysis by
turning on, in addition, world volume gauge fields on the rotating probe D7-brane.
In Sec.\,4, we first derive the induced  world volume metric and compute the 
Hawking temperature on the rotating probe in the presence of the world volume
electric field. We then derive the energy--stress tensor on the rotating brane
and compute its backreaction and energy dissipation. In Sec.\,5, we first derive
the induced world volume metric and compute the Hawking temperature on the rotating
probe in the presence of the world volume magnetic field. We then derive the 
energy--stress tensor on the rotating brane and compute its backreaction and energy
dissipation. In Sec.\,6, we discuss our results and summarize with future outlook.

\section{Review of the Kuperstein-Sonnenschein model}

The specific ten-dimensional background that we would like to consider
is the KW solution, \cite{Klebanov:1998hh}, obtained from taking the 
near horizon limit of a stak of $N$ background D3-branes on the conifold
point.
The conifold is defined by a $2\times 2$ matrix $W$ in $\mathbb{C}^4$, 
and by a radial coordinate, given by (for more details see 
refs.\,\cite{Evslin:2007ux} and ref.\,\cite{Kuperstein:2008cq}):

\begin{equation}
\label{condef}
\det W=0,\;\;\;\;\ r^2=\frac{1}{2}\text{tr}(W\,W^{\dagger}).
\end{equation}
Here the radial coordinate is fixed by the virtue of the scale invariance
of the determinantal equation defining the conifold. This equation describes
a cone over a five-dimensional base. The matrix $W$ in (\ref{condef}) is 
singular and may be represented by:

\begin{eqnarray}
\label{WCon}
W&=&\sqrt{2}r uv^{\dagger},\;\;\;\;\ u u^{\dagger}=v v^{\dagger}=1.
\end{eqnarray}
This representation is not unique, since it is invariant under 
$(u,v)\rightarrow \exp(i\varphi)(u,v)$. Nonetheless, one can use
$u$ and $v$ and define a matrix $X\in SU(2)$ by $u=X v$ and obtain
a unique solution $X= u v^{\dagger}-\epsilon u v^{T}\epsilon$. The
matrix $X$ is invariant under the exponential map in (\ref{WCon})
and therefore describes an $S^3$. On contrary, $v$ is yet defined
modulo $v=\exp(i\varphi)v$ and therefore describes an $S^2$. Given
$r$, $X$ and $v$ one can set $u$ and thereby $W$, and get 
$W=\sqrt{2} r X v v^{\dagger}$. Thus the base of the conifold,
denoted $T^{1,1}$, where $r=\text{const.}$, is uniquely parameterized
by $X$ and $v$ and hence the topology of $T^{1,1}$ is identified with
$S^3\times S^2$. In addition, we note that the product $v v^{\dagger}$
is hermitian with eigenvalues 1 and 0 and hence can be written in terms of
an $SU(2)$ matrix $\gamma$ (and  $\gamma^{\dagger}$), set by $v$ up to a
gauge  transformation $\gamma\rightarrow \gamma\exp(i\varphi\sigma_3)$.
As for $v$, $\gamma$ defines an $S^2$ and by a gauge
transformation one can always write $\gamma=\exp(i\varphi\sigma_3)
\exp(i\theta\sigma_2)$.

Placing $N$ regular D3-branes on the conifold backreacts on the geometry,
\cite{Klebanov:1998hh}, produces the 10D warped line element in terms of
$S^3\times S^2$ coordinates as \cite{Kuperstein:2008cq}:

\begin{eqnarray}
\label{10DKWmet}
ds_{10}^2&=&h(r)^{-1/2}dx_{n}dx^{n}+h(r)^{1/2}(dr^2+r^2 ds_{T^{1,1}}^2),\\
\label{6DConmet}
ds_{T^{1,1}}^2&=&\frac{r^2}{3}\bigg[\frac{1}{4}(\Omega_1^2+\Omega_2^2)
+\frac{1}{3}\Omega_3^2+\Big(d\theta-\frac{1}{2}\Omega_2\Big)^2+
\Big(\sin\theta d\phi-\frac{1}{2}\Omega_1\Big)^2\bigg],
\end{eqnarray}
with the warp factor

\begin{equation}
h(r)=\frac{L^4}{r^4},\;\;\;\;\ \text{and}\;\;\;\ L^4\equiv \frac{27\pi}{4}
g_sN(\alpha^{\prime})^2.
\end{equation}
Here the first term in (\ref{10DKWmet}) is the usual four-dimensional 
Minkowski spacetime metric and the second term is the metric on a 
six-dimensional Ricci-flat cone, the Calabi-Yau cone, given by the conifold
metric (\ref{6DConmet}) \cite{Candelas:1989js,Evslin:2007ux}. In this metric,
the radial coordinate, $r$, is defined by (\ref{condef}), $\theta$ and $\phi$
parameterize the $S^2$, and $\Omega_i$ are one-forms parameterizing the $S^3$.
The $\Omega_i$s can be represented by Maurer-Cartan one-forms $w_i$ via two 
$SO(3)$ matrices (we do not write them out here) parameterized by $\theta$
and $\phi$, respectively,  which show that the $S^3$ is fibered trivially over
the $S^2$. In order to have a valid supergravity solution, (\ref{10DKWmet}),
the number of D3-branes, $N$, placed on the conifold has to be  large, and the
string coupling, $g_s$, has to be small, so that $g_sN\gg1$; 
$\alpha^{\prime}=l_s^2$ denotes the string scale. Here the dilaton is constant,
and the other non-trivial background field is a self-dual R--R five-form flux of
the form:

\begin{equation}
 F_5=\left(\frac{4r^3}{g_sL^4}\right)dr\wedge dt\wedge dx\wedge dy \wedge dz
 -\left(\frac{L^4}{27g_s}\right) \sin\theta d\theta\wedge d\phi\wedge\Omega_1
 \wedge\Omega_2\wedge\Omega_3.
\end{equation}

The above supergravity solution, $adS_5\times T^{1,1}$, is dual to  $\mathcal{N}=1$
superconformal field theory with the gauge group $SU(N)\times SU(N)$ coupled
to two chiral superfields, $A_i$, in the $(\mathbf{N},\mathbf{\overline{N}})$
representation and two chiral superfields, $B_j$, in the $(\mathbf{\overline{N}}
,\mathbf{N})$ representation of the gauge group. The fields $A_i$ and $B_j$, and
so the gauge group factors $SU(N)$, get interchanged by the $\mathbb{Z}_2$--symmetry
of the conifold geometry, acting as $W\rightarrow W^{T}$, with $W$ given by (\ref{WCon}).
The formula $W=\sqrt{2}r X v v^{\dagger}$, however, shows that by $\mathbb{Z}_2$--symmetry
$(X,v) \rightarrow (X^{T}, (Xv)^{*})$. Thus, in the flavor brane embedding configuration
reviewed next, the $\mathbb{Z}_2$--symmetry gets broken. This is because the unit vector
$v$ parameterizes the $S^2$ where the position of the flavor brane depends only on the 
radial coordinate of the conifold and is independent from $X$, so not respecting the
$\mathbb{Z}_2$ transformation of $v$. In addition, the fields $A_i$ and $B_j$ transform
as a doublet of the first and as a singlet of the second factor in the 
$SU(2)_1\times SU(2)_2$--symmetry, acting as $W\rightarrow S_1 W S_2^{\dagger}$, with 
$W$ given by (\ref{WCon}) and $S_{1,2}$ denoting $SU(2)$ matrices. The formula 
$W=\sqrt{2}r X v v^{\dagger}$ shows that  $(X,v)\rightarrow (S_1 XS_2^{\dagger}, S_2 v)$.
Thus, in the flavor brane  embedding reviewed next, the $S_2$ gets broken, while the $S_1$
is preserved.

We now would like to consider the brane embedding configuration of 
ref.\,\cite{Kuperstein:2008cq}, embedding probe D7-branes into the
KW background, corresponding to adding flavored quarks to its
dual gauge theory. In the KW, the D7-brane spans the spacetime and
radial coordinates $\{t,x_i,r\}$ ($i=1,2,3$) of $adS_5$ in the 
01234-directions, and the three-sphere $S^3$ of $T^{1,1}$ parameterized
by the forms $\{\Omega_i\}$ in the 567-directions. Thus the transversal
space consists of the two-sphere $S^2$ of $T^{1,1}$ parameterized by the 
coordinates $\theta$ and $\phi$ in the 89-directions. This is represented
by the array:

\[ \begin{array}{lcr}
\;\;\;\;\,\  0\,\,\, 1\,\,\, 2\,\,\; 3\,\,\, 4\,\,\, 5\,\,\, 6\,\,\, 7\,\,\, 8\,\, 9\\
\mbox{D3}  \times \times\times\times\\
\mbox{D7}  \times \times \times \times\times\times\times\times \\ 
\end{array}\]
Here we note that, as $w_i$ are left-invariant forms, the ansatz preserves one
of the $SU(2)$ factors of the global symmetry group of the conifold, 
$SU(2)\times SU(2)\times U(1)$. Thus, one may assume that the coordinates $\theta$
and $\phi$ are independent of the $S^3$ coordinates. The embedding breaks one of
$SU(2)$, but by expanding the action around the solution it can be shown that 
contribution from the nontrivial $S^3$ show up only at the second order fluctuations.
Thus, one can assume, in classical sense, that $\theta$ and $\phi$ depend only on the
radial coordinate, $r$, of the conifold geometry.

To write down the action of the D7-brane, we note that the KW solution
contains only the R--R four-form fluxes and therefore the Chern-Simons
part does not contribute. Thus, the action of the D7-brane is simply 
given by the DBI action as:

\begin{equation}
\label{AC}
S_{D7}=-T_{D7}\int{d^8\xi\sqrt{-det(g_{ab}^{D7}+(2\pi\alpha^{\prime})F_{ab})}}
\;\;\;\; \text{with}\;\;\;\;\; T_{D7}=\frac{1}{(2\pi)^7 g_s(\alpha^{\prime})^4 }.
\end{equation}
Here $T_{D7}$ is the tension, $\xi^a$ are the world-volume
coordinates, $g_{ab}^{D7}$ is the induced world-volume metric, and $F_{ab}$ is
the $U(1)$ world-volume field strength.

For the embedding of the D7-branes, we note that there are two choices.
One choice is to place the D7-branes on two separate points on the $S^2$
and stretch them down to the tip of the conifold at $r=0$ where the $S^2$
and $S^3$ shrink to zero size. This is called \emph{V}-shape configuration.
The other choice is to place the D7-branes on the $S^2$ and smoothly merge
them into a single stack somewhere at $r=r_0$ above the tip of the cone. 
This is called \emph{U}-shape configuration. In both of the \emph{U}-shape
and \emph{V}-shape configurations, the D7-brane(s) wrap the $adS_5$ and the
$S^3$, as in the array above. However,  on the transversal $S^2$, there are
two different pictures. In the \emph{V}-shape configuration, the D7-branes
appear as two separate fixed points whereas the \emph{U}-shape configuration
produces an arc along the equator. The position of the two points, giving 
the position of the D7-branes, depends on $r$ and they smoothly connect in
the midpoint arc at $r=r_0$. In supergravity, this configuration produces
a 1-parameter family of D-brane profiles with the parameter $r_0$ giving
the minimal radial extension of the D7-brane.

By the choice of the world volume fields $\phi=\phi(r)$ and $\theta=\theta(r)$,
it is easy to derive the induced world volume metric and obtain from (\ref{AC})
the action of the form:
\begin{equation}
\label{KSA}
S_{D7}=-\tilde{T}_{D7}\int{drdt\,r^3\sqrt{1+\frac{r^2}{6}({\theta^{\prime}}^2+
\sin^2\theta{\phi^{\prime}}^2)}},
\end{equation}
where $\tilde{T}_{D7}=N_fV_{\mathbb{R}^3}V_{\mathbb{S}^3}T_{D7}$.
The Lagrangian in (\ref{KSA}) is $SU(2)$ invariant and therefore one can
restrict motion to the equator of the $S^2$ parameterized by $\phi$ and
$\theta=\pi/2$. 

The solution of the equation of motion from the action (\ref{KSA}) yields
a one-parameter family of D7-brane profiles of the form:

\begin{eqnarray}
\label{KUPSol.}
\phi(r)=\sqrt{6}r_0^4\int_{r_0}^{r}{\frac{dr}{r\sqrt{r^8-r_0^8}}}=
\frac{\sqrt{6}}{4}\cos^{-1}\bigg(\frac{r_0}{r}\bigg)^4.
\end{eqnarray}
At $r_0=0$, the solution (\ref{KUPSol.}) describes two separate branches,
a disconnected D7 and an anti D7-brane pair, and the configuration is of
\emph{V}-shape. At $r_0>0$, the two branches merge at $r=r_0$ and the
configuration is of \emph{U}-shape. We also note that when the configuration
is \emph{U}-like, taking the limit $r\rightarrow r_0$ implies 
$\phi^{\prime}(r)\rightarrow\infty$.

The above solution has a number of important features. First, in the \emph{V}-shape
configuration one can see from  $d\theta=d\phi=0$ that the induced world volume
metric is that of $adS_5\times S^3$ and the configuration describes the conformal
and chiral symmetric phase. On contrary, in the \emph{U}-shape configuration the
induced world volume metric has no $adS$ factor and the conformal and chiral flavor
symmetry of the dual gauge theory must be broken spontaneously. Second, the 
asymptotic UV limit, $r\rightarrow\infty$, is described by two constant solutions
$\phi_{\pm}=\pm\sqrt{6}\pi/8$, giving an asymptotic UV separation between the branes,
and an asymptotic expansion, respectively, as:

\begin{equation}
\Delta\phi=\phi_{+}-\phi_{-}=\frac{\sqrt{6}\pi}{4}, \;\;\;\;\;\
\phi\simeq\pm\frac{\sqrt{6}\pi}{8}\pm\frac{\sqrt{6}}{4}
\left(\frac{r_0}{r}\right)^4+\cdots.
\end{equation}
The asymptotic UV separation between the branes is independent from $r_0$.
In the dual gauge field theory $r_0$ corresponds to a normalizable mode, a
vacuum expectation value (VEV). The fact that $r_0$ is a modulus, or a flat
direction, implies the spontaneous breaking of the conformal symmetry. The
expansion shows that a $\Delta=4$ marginal operator has a VEV fixed by $r_0$
as:

\begin{equation}
\left< O\right>\sim \frac{r_0^4}{
(\alpha^{\prime})^2},
\end{equation}
with its fluctuations giving the Goldstone boson associated with the conformal
symmetry breakdown. Third, the solutions $\phi_{\pm}$, giving an $r_0$-independent
UV separation, make the brane anti-brane interpretation natural. This is because
the brane worldvolume admits two opposite orientations once the asymptotic points 
$\phi_{\pm}$ are approached. Fourth, the presence of both the D7 and anti D7-brane
guarantees tadpole cancellation and annihilation of total charge on the transverse
$S^2$, and it breaks supersymmetry explicitly with the embedding being non-holomorphic.

To this end, one also notes that the above setup cannot be embedded in
the $adS_5\times S^5$ solution. This because the $S^5$ contains no 
nontrivial cycle and therefore the D7-brane will shrink to a point on
the $S^5$. This problem may be fixed by a specific choice of boundary
conditions at infinity, but this turns out to be incompatible with the
\emph{U}-shape configuration of interest. In addition, tadpole cancellation
by an anti-D7-brane is not required, since one has no 2-cycle as in the
conifold framework.

\section{Rotating D7-branes in $adS_5\times T^{1,1}$} 

\subsection{Induced metric and temperature}

We now would like to consider the D7-brane configuration in $adS_5\times T^{1,1}$
reviewed in the previous section, including additional spin degrees of freedom.
By spherical symmetry, we let in our setup the D7-brane rotate in the $\phi$
direction of the $S^2$ with conserved angular momentum. Therefore in our analysis
we may let $\phi$ depend on time as well, so that $\dot{\phi}(r,t)=\omega=const.$,
where $\omega$ denotes the angular velocity. This will allow us to construct rotating
solutions. So, henceforth we consider the world-volume fields in the action (\ref{AC})
to be $\theta(r)$ and $\phi(r,t)$. In the next sections, we will also consider cases
including, in addition, the contribution world-volume fields strengths, $F_{ab}$, in
(\ref{AC}), corresponding to world-volume electric and magnetic fields.

Thus, we will consider an ansatz for the D7-brane world volume fields $\theta=\theta(r)$,
$\phi(r,t)=\omega t+f(r)$, and for now $F_{ab}=0$. With this ansatz, it is straightforward
to find the components of the induced world volume metric on the D7-brane, $g_{ab}^{D7}$,
and compute the determinant in (\ref{AC}), giving the DBI action as:

\begin{eqnarray}
\label{dbiac1}
S_{D7}=-\tilde{T}_{D7}\int{dr dt\, r^3\sqrt{1-\frac{L^4\dot{\phi}^2}{6r^2}+\frac{r^2}{6}
\left({\theta^{\prime}}^2+{\phi^{\prime}}^2\sin^2\theta\right)-\frac{L^4}{18}
\sin^2\theta{\theta^{\prime}}^2\dot{\phi}^2}}.
\end{eqnarray} 
Here we note that by setting $\dot{\phi}=\omega=0$, our action (\ref{dbiac1}) reduces
to that of the Kuperstein--Sonnenschein model, (\ref{KSA}).
As in the Kuperstein--Sonnenschein model reviewed in Sec.\,2, we set $\theta=\pi/2$ and
restrict brane motion to the equator of the $S^2$ sphere. Thus, in our set up we let, in
addition, the probe rotate about the equator of the $S^2$. The equation of motion from
the action (\ref{dbiac1}) then take the form:

\begin{equation}
\label{D7eq}
\frac{\partial}{\partial r}\Bigg[\frac{r^5\phi^{\prime}}{\sqrt{1+\frac{r^2(\phi^{\prime})^2}{6}
-\frac{L^4\dot{\phi}^2}{6r^2}}}\Bigg]=
\frac{\partial}{\partial t}\Bigg[\frac{r\dot{\phi}}{\sqrt{1+\frac{r^2(\phi^{\prime})^2}{6}
-\frac{L^4\dot{\phi}^2}{6r^2}}}\Bigg].
\end{equation}

\begin{figure}[t]
\begin{center}
\epsfig{file=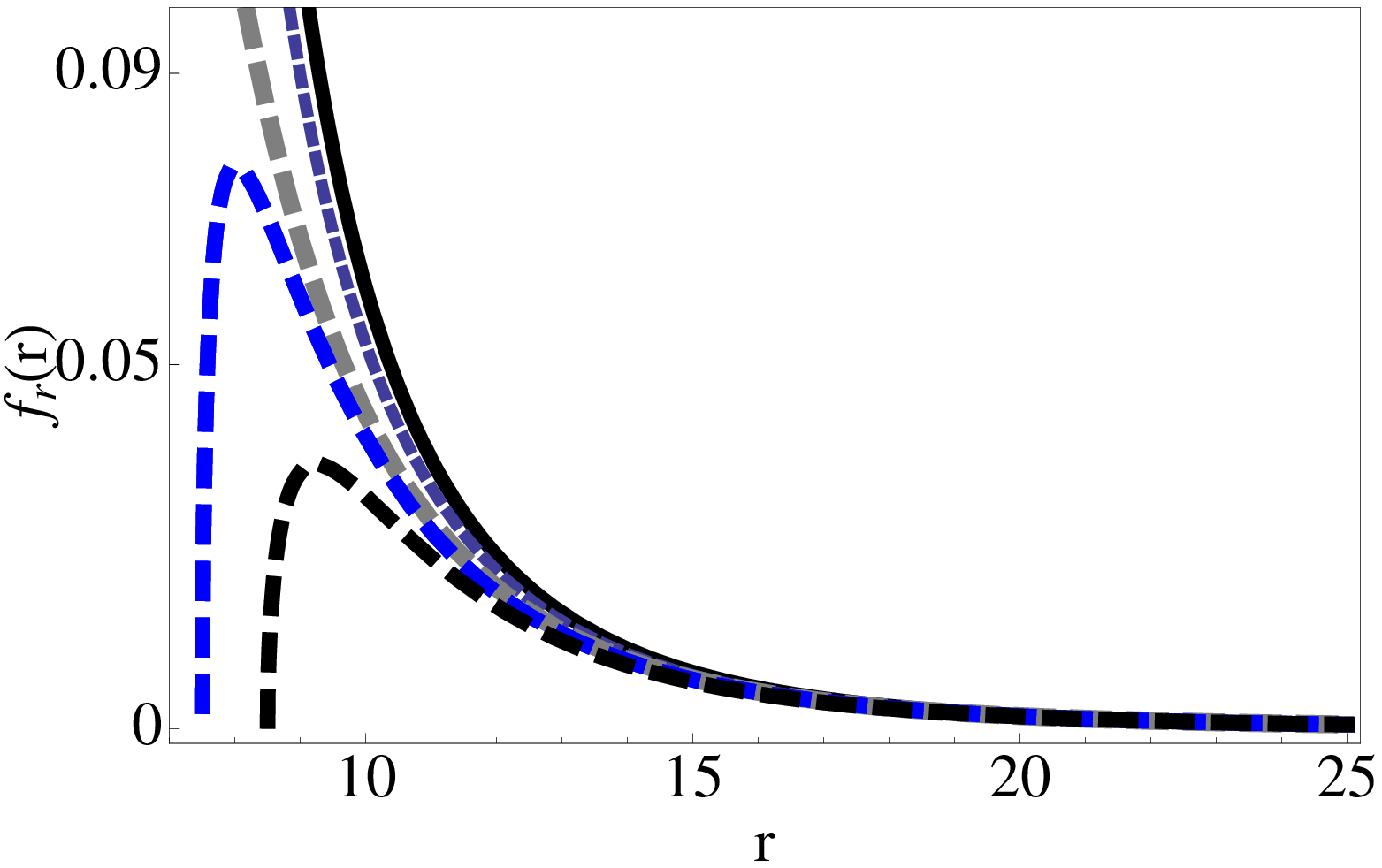,width=.50\textwidth}~~\nobreak
\epsfig{file=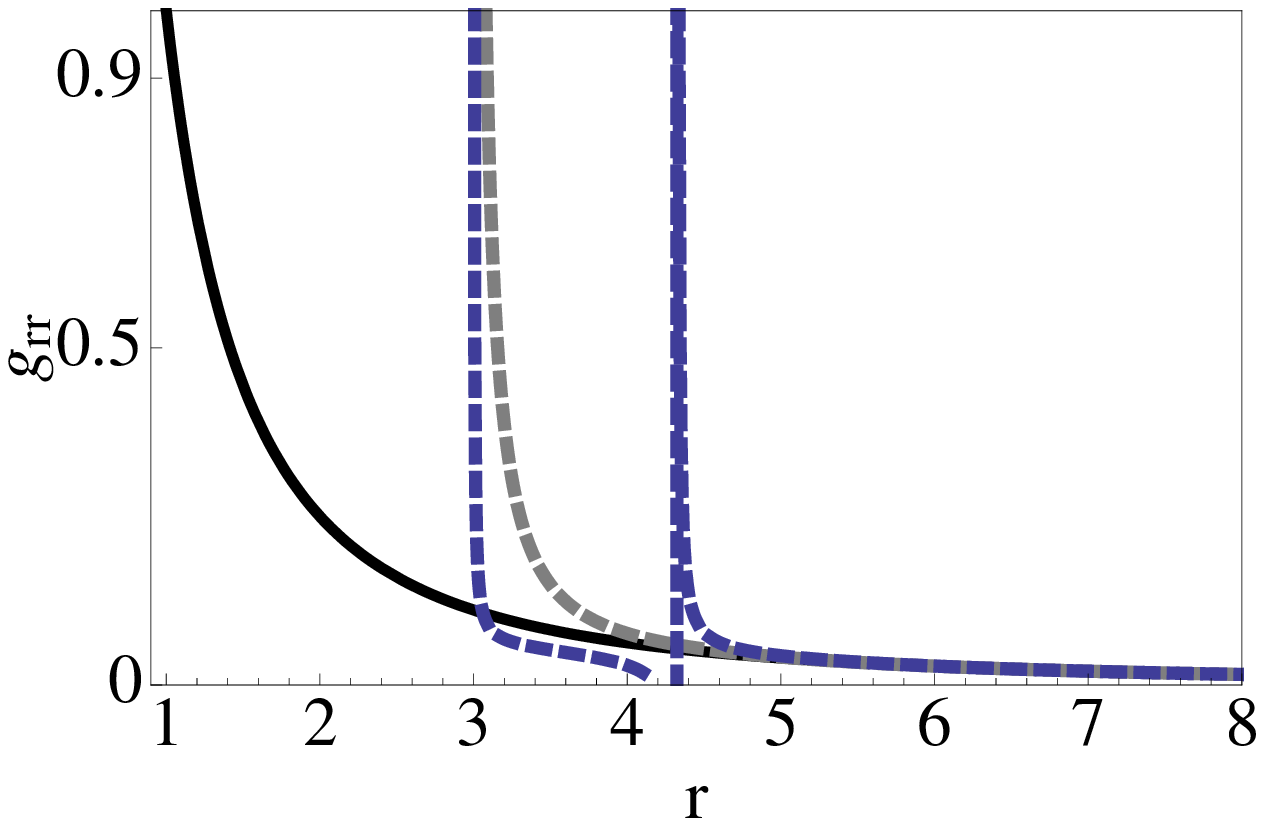,width=.50\textwidth}
\caption{[Left] The behavior of the derivative of the world
volume field with respect to $r$ with $L=1, r_0=7$, $\omega=0$
(black-solid), $\omega=5$ (dark blue-dashed), $\omega=7$ 
(gray-dashed), $\omega=7.5$ (blue-dashed), and $\omega=8.5$
(black-dashed). [Right] The behavior of the $g_{rr}$ component
of the induced world volume metric with $L=1,r_0=0, \omega=0$
(black-solid), $r_0=3, \omega=0$ (gray-dashed), and 
$r_0=3,\omega=7.5$ (blue-dashed).}
\label{fig:KWgp}
\end{center}
\end{figure}

Consider rotating solutions of the form:

\begin{eqnarray}
\label{rotsol}
\phi(r,t)&=&\omega t+f(r),\;\;\;\;\ f(r)=\sqrt{6}r_0^4\int_{r_0}^{r}{\frac{dr}{r}
\sqrt{\frac{1-L^4\,\overline{\omega}^2/r^2}{r^8-r_0^8}}}.
\end{eqnarray}
Here we note that by setting $\overline{\omega}=\omega/\sqrt{6}=0$, our solution
(\ref{rotsol}) reduces to that of the Kuperstein--Sonnenschein model, 
\cite{Kuperstein:2008cq}, reviewed in Sec.\,2 (see Eq.\,(\ref{KUPSol.})). It is
also clear that the above rotating solution has two free parameters, the angular
velocity $\omega$ and the minimal radial extension $r_0$. The solution (\ref{rotsol})
describes brane motion with spin starting and ending up at the boundary. The brane
comes down from the UV boundary at infinity, bends at the minimal extension in the IR,
and backs up the boundary. We also note that when $r$ is large, the behavior of the
derivative of $f(r)$ with respect to $r$, denoted $f_r(r)$, with and without $\omega$
is the same (see Fig.\,\ref{fig:KWgp}). This shows that in such limit the derivative
of $f_r(r)$ integrates to the $\phi(r)$ of the Kuperstein--Sonnenschein model 
(see Sec.\,2) with the boundary values $\phi_{\pm}$ in the asymptotic UV limit, 
$r\rightarrow\infty$ (see also Fig.\,\ref{fig:KWgp}). However, we note that in the
(opposite) IR limit, i.e., when $r$ is small, the behavior of $f_r(r)$ does depend on
$\omega$. Inspection of (\ref{rotsol}) shows that in the IR only for certain values of
$\omega$ the behavior of $f_r(r)$ with compares to that of without $\omega$ 
(see Fig.\,\ref{fig:KWgp}). This shows that in the IR and within specific range of
$\omega>0$ the behavior of $f_r(r)$ (here) compares to that of $\phi^{\prime}(r)$ in
the Kuperstein--Sonnenschein model (see Sec.\,2), where  $\phi^{\prime}(r)\rightarrow\infty$
in the IR limit $r\rightarrow r_0$, consistent with U-like embedding.

To derive the induced metric on the D7-brane, we put the rotating solution (\ref{rotsol})
into the background metric (\ref{10DKWmet}) and obtain:

\begin{eqnarray}
\label{ind1}
ds_{ind.}^2&=&-\frac{(3r^2-L^4\omega^2)}{3L^2}dt^2+\frac{L^2}{r^2}\left[\frac{3r^2(r^8-r_0^8)
+r_0^8(6r^2-L^4\omega^2)}{3r^2(r^8-r_0^8)}\right]dr^2\notag\\ &&+\frac{2L^2 \omega\, r_0^4}
{3r^2}\sqrt{\frac{6r^2-L^4\,\omega^2}{r^8-r_0^8}}drdt+\frac{r^2}{L^2}(dx^2+dy^2+dz^2)\notag\\
&&+\frac{L^2}{3}\left[\frac{1}{2}(\Omega_1^2+\Omega_2^2)+\frac{1}{3}\Omega_3^2+ \omega\Omega_1
dt- \frac{r_0^4}{r^2}\sqrt{\frac{6r^2-L^4\,\omega^2}{r^8-r_0^8}}\Omega_1 dr\right].\notag\\
\end{eqnarray}
Here we note that by setting $\omega=0$, our induced world volume
metric (\ref{ind1}) reduces to that of the Kuperstein--Sonnenschein
model, \cite{Kuperstein:2008cq}, reviewed in Sec.\,2. In this case,
for $r_0=0$ the induced world volume metric is that of $adS_5\times S^3$
and the dual gauge theory describes the conformal and chiral symmetric
phase. On contrary, for $r_0>0$ the induced world volume metric has no
$adS$ factor and the conformal invariance of the dual gauge theory must
be broken in such case. In order to find the world volume horizon and
Hawking temperature, we first eliminate the relevant
cross term. To eliminate the relevant cross-term, we consider a coordinate
transformation:

\begin{equation}
\tau=t-\omega\,L^4 r_0^4 \int{\frac{dr\,(6r^2-L^4\,\omega^2)^{1/2}}
{r^2(3r^2-L^4 \omega^2)(r^8-r_0^8)^{1/2}}}.
\end{equation}
The induced metric on the rotating D7-brane(s) then takes the form:

\begin{eqnarray}
\label{ind21}
ds_{ind.}^2 &=&-\frac{(3r^2-L^4 \omega^2)}{3L^2}d\tau^2+\frac{L^2}{r^2}
\left[\frac{(3r^2-L^4 \omega^2)(r^8-r_0^8)+r_0^8(6r^2-L^4 \omega^2)}{(3r^2
-L^4 \omega^2)(r^8-r_0^8)}\right]dr^2\notag\\ && +\frac{r^2}{L^2}(dx^2+
dy^2+dz^2)\notag\\ && +\frac{L^2}{3}\left[\frac{1}{2}(\Omega_1^2+\Omega_2^2)
+\frac{1}{3}\Omega_3^2-\omega\Omega_1d\tau-\frac{3\,r_0^4}{3r^2-L^4 \omega^2}
\sqrt{\frac{6r^2-L^4\,\omega^2}{r^8-r_0^8}}\Omega_1 dr\right].\notag\\
\end{eqnarray}
Here we note that for $r_0=0$ the induced world volume metric (\ref{ind21}) has no
horizon, such that $-g_{tt}=g^{rr}=0$, and therefore not given by the black hole
geometry. On contrary, for $r_0>0$ the induced world volume horizon is described
by the the horizon equation of the form:

\begin{equation}
\label{HorEq1}
H(r)=r^2(r^8-r_0^8)(3r^2-L^4\omega^2)=0.
\end{equation}
This equation has two obvious real positive definite zeros
(see also Fig.\,\ref{fig:KWgp}). The thermal horizon of the
induced world volume black hole geometry can be identified
with the solution of the horizon equation (\ref{HorEq1})
as $-g_{tt}=g^{rr}=H(r)=3r_h^2-L^4\omega^2=0$. It is clear
that the world volume horizon increases/decreases with 
increasing/decreasing the angular velocity $\omega$, as it 
should. It is also clear from the horizon equation 
(\ref{HorEq1}) that the horizon must grow from the minimal
extension $r_0\neq0$ with increasing the angular velocity 
$\omega$. We therefore conclude at this point that when $r_0$
is positive definite ($r_0>0$) and spin is turned on ($\omega>0$),
the induced world volume metric on the rotating probe has a
thermal horizon growing with increasing the
angular velocity.

The Hawking temperature can be found from this induced metric in the form:

\begin{eqnarray}
\label{TH1}
T &=& \frac{(g^{rr})^{\prime}}{4\pi}\bigg|_{r=r_h}=\frac{3r_h^3(r_h^8-r_0^8)}
{2\pi L^2 r_0^8(6r_h^2-L^4\omega^2)}=\frac{r_h (r_h^8-r_0^8)}{2 \pi\,L^2 r_0^8}.
\end{eqnarray}
Here, as before, $r_0$ denotes the minimal radial extension. It is clear from
(\ref{TH1}) that at $r_h=r_0$ the temperature of the world volume black hole 
solution is precisely zero, $T=0$. It is also clear from (\ref{TH1}) that 
the temperature increases with growing horizon size, $r_h>r_0$. Inspection of
(\ref{TH1}) also shows that the temperature of the configuration increases with
decreasing the modulus $r_0$ (see Fig.\,\ref{fig:KWp1}). It is also clear 
(from Fig.\,\ref{fig:KWp1}) that when $r_0$ is decreased, at sufficiently large
horizon size the configuration admits high temperatures. Furthermore, decreasing
the value of $r_0$ further shows a large separation between the temperatures, 
$T_{r_0<1}/T_{r_0>1}\simeq 10^8$ (see Fig.\,\ref{fig:KWp1}). We may therefore
conclude that when the size of the modulus $r_0$ is decreased, whereby the chiral
and conformal symmetric phase is approached, the temperature increases dramatically;
however, as discussed above, we note that at $r_0=0$ the induced world volume metric
(\ref{ind21}) has no thermal horizon and Hawking temperature. 

We also note that by considering the backreaction of this solution to the SUGRA
background, given by the KW solution $adS_5\times T^{1,1}$, one naturally expects
the  D7-brane to form a very small black hole in KW, describing a locally thermal
gauge field theory in the probe limit. Accordingly, the rotating D7-brane describes
a thermal object in the dual gauge field theory. In the KW example here, the system
is dual to  $\mathcal{N}=1$ gauge theory coupled to a quark. Since the gauge theory
itself is at zero temperature while the quark is at finite temperature $T$, given by
(\ref{TH1}), the system is in non-equilibrium steady state. However, as discussed
blow, the energy from the flavor sector will eventually dissipate to the gauge theory.

In the above analysis, the backreaction of the D7-brane to the
supergravity background has been neglected since we considered
the probe limit. It is instructive to see to what extend this
can be justified. We note that the components of the stress--energy
tensor of the D7-brane take the form\footnote{Here we are using the
energy-stress tensor defined by $J_{N}^{M}=\frac{2}{\sqrt{-g}}
\frac{\delta S}{\delta g_{M L}}g_{L N}$, where $g_{MN}$ denotes
the bulk metric with $M$ and $N$ running over all ten coordinates
of the ten-dimensional spacetime. This satisfies the equation of
motion $\nabla_M J_N^M=0$. For static spacetime, this reduces to
$\partial_M(J_t^M\sqrt{-g})=0$, which leads to the energy 
$E=\int{dr\sqrt{-g}J_{t}^{t}}$.}:

\begin{figure}[t]
\begin{center}
\epsfig{file=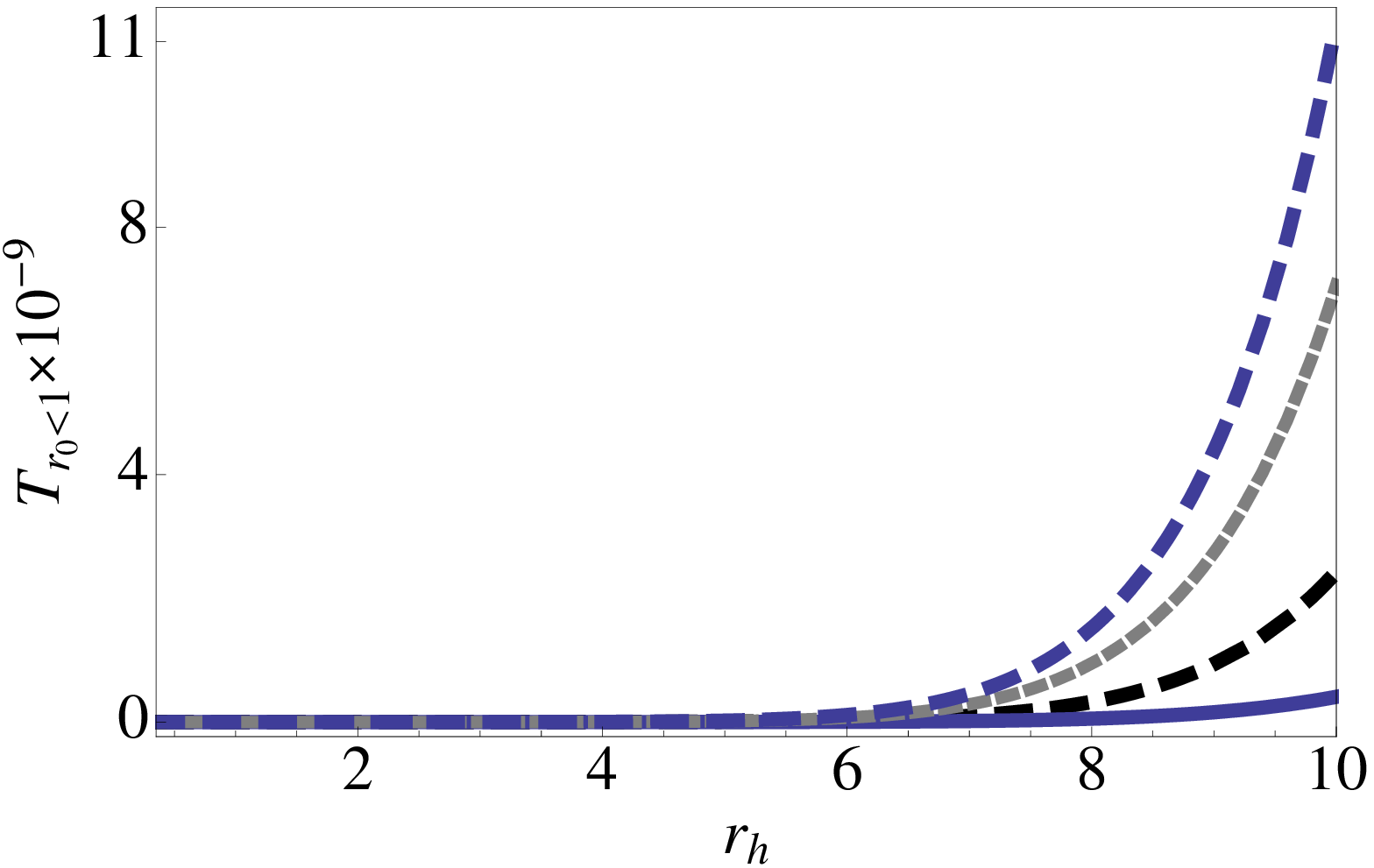,width=.5\textwidth}~~\nobreak
\epsfig{file=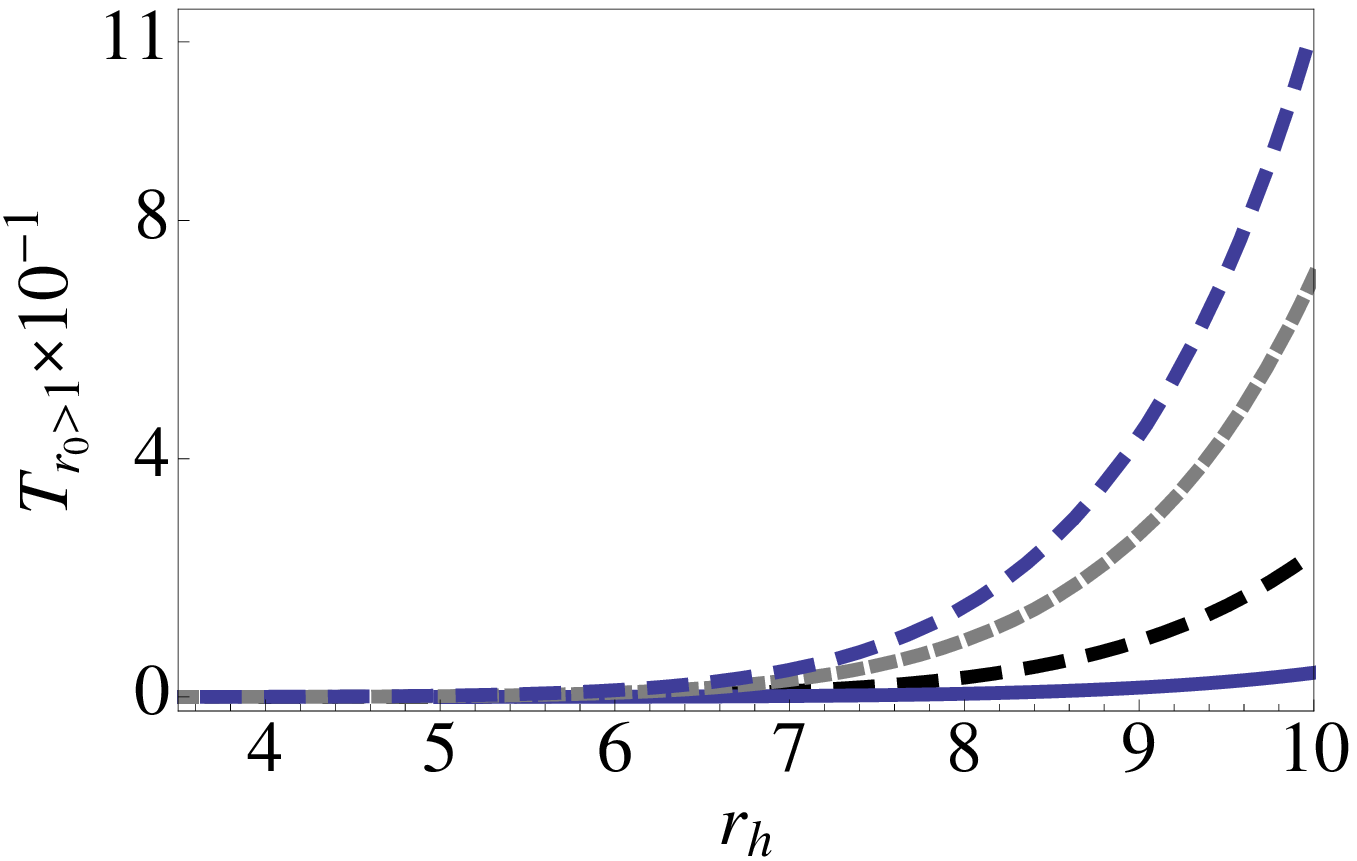,width=.5\textwidth}
\epsfig{file=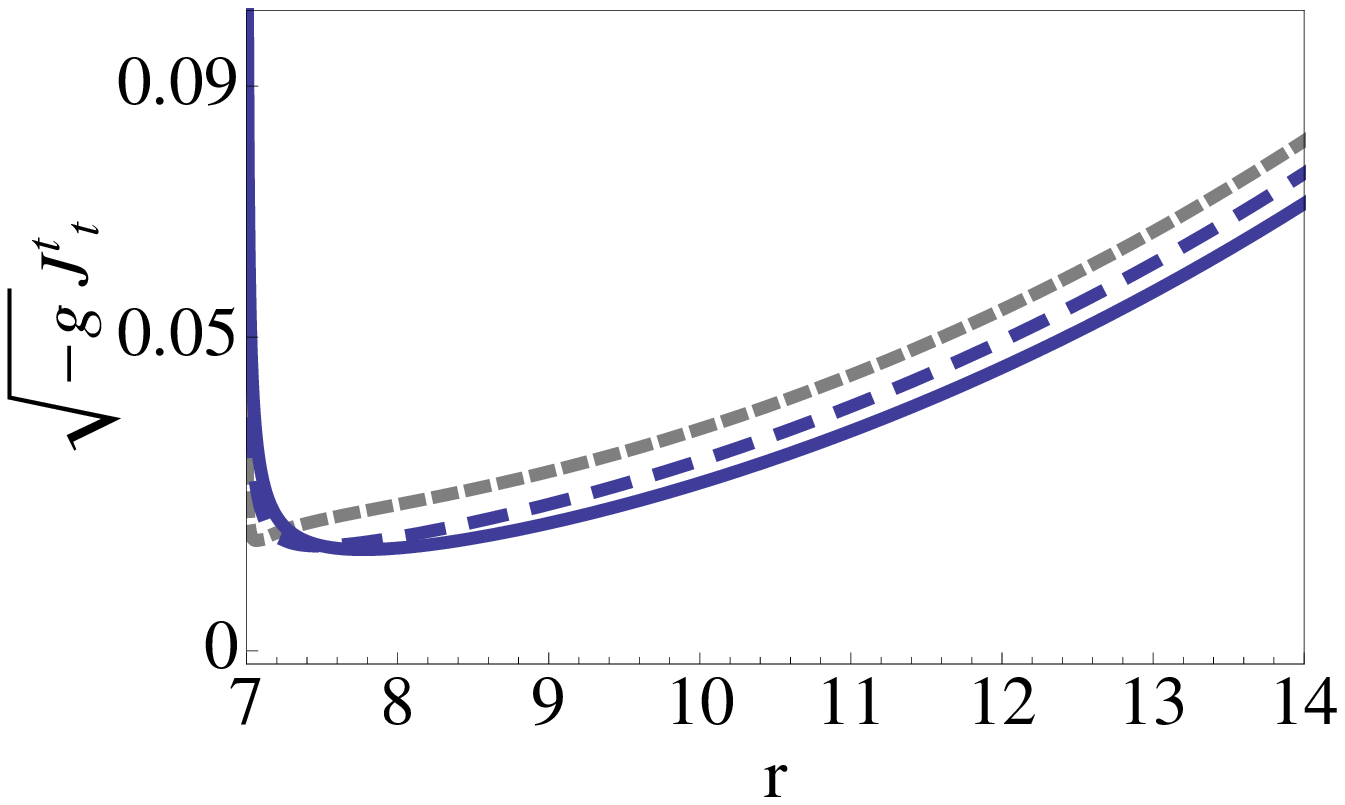,width=.5\textwidth}~~\nobreak
\epsfig{file=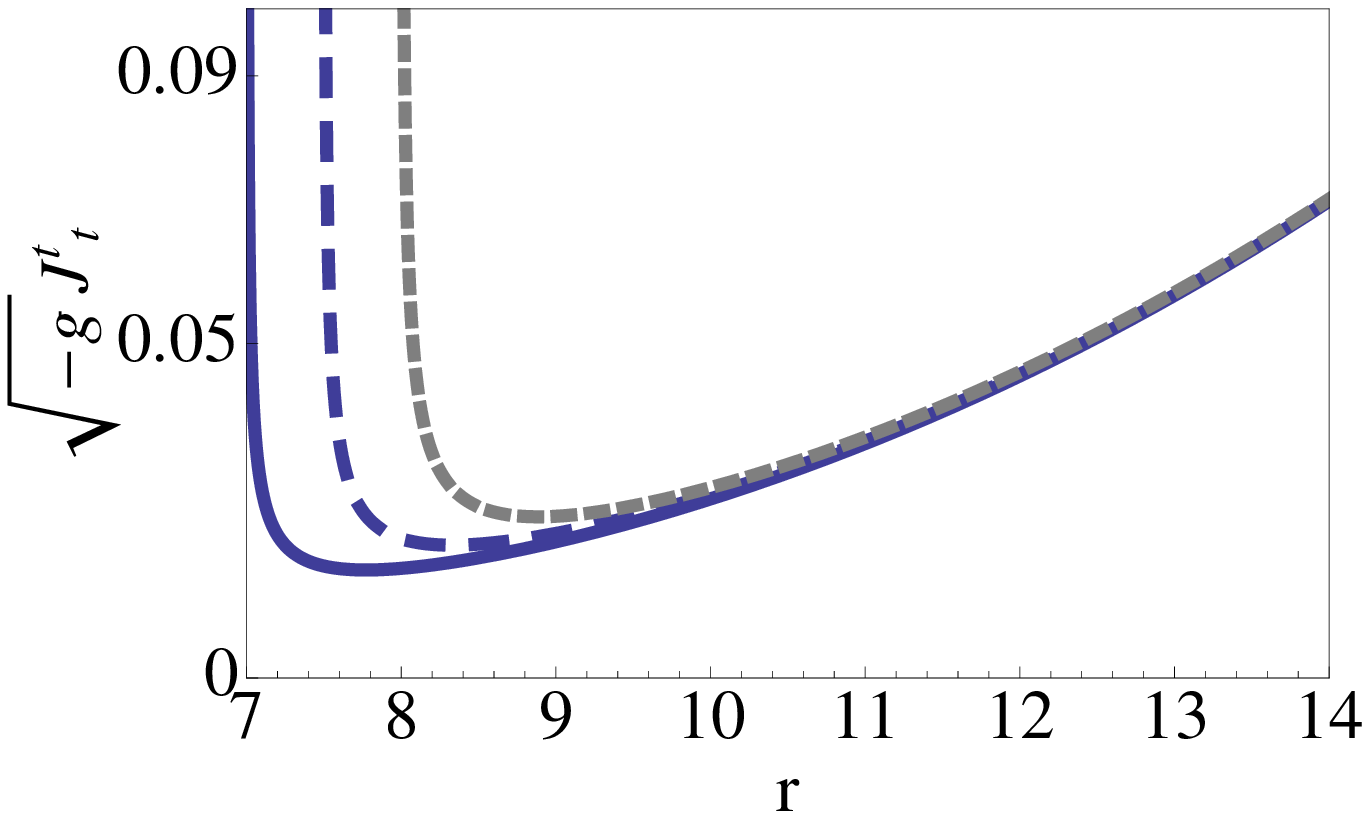,width=.5\textwidth}
\caption{[Up-Left] The behavior of the temperature with $r_0=5$
(solid), $r_0=4$ (black-dashed), $r_0=3.5$ (gray-dashed), $r_0=3.3$
(blue-dashed), and $L=10$. [Up-Right] The behavior of the temperature
with $r_0=0.5$ (solid), $r_0=0.4$ (black-dashed), $r_0=0.35$ 
(gray-dashed), $r_0=0.33$ (blue-dashed), and $L=10$. [Down-Left]
The behavior of the energy-density, $\sqrt{-g}J_t^t$, with $\omega=6.8$
(gray-dashed), $\omega=5$ (dashed), $\omega=1$ (solid), and $r_0=7$.
[Down-Right] The behavior of the energy-density with $r_0=8$ 
(gray-dashed), $r_0=7.5$ (dashed), $r_0=7$ (solid), and $\omega=1$.
In all cases, $L=\alpha^{\prime}=1$ and $g_s=0.1$.}
\label{fig:KWp1}
\end{center}
\end{figure}

\begin{eqnarray}
\label{T1}
\sqrt{-g}J_{t}^{t}&\equiv&\frac{\tilde{T}_{D7}\,r^3(1+r^2(\phi^{\prime})^2/6)}
{\sqrt{1+r^2(\phi^{\prime})^2/6-L^4\dot{\phi}^2/6r^2}}=
\frac{\tilde{T}_{D7}(r^{10}-r_0^8L^4\overline{\omega}^2)}{r^2\sqrt{(r^8-r_0^8)
(r^2-L^4\overline{\omega}^2)}},\\ \label{T2}\sqrt{-g}J_{r}^{r}&\equiv&
-\frac{\tilde{T}_{D7}\,r^3(1-L^4\dot{\phi}^2/6r^2)}{\sqrt{1+r^2(\phi^{\prime})^2/6
-L^4\dot{\phi}^2/6r^2}}=-\frac{\tilde{T}_{D7}}{r^2}\sqrt{(r^8-r_0^8)(r^2-L^4
\overline{\omega}^2)},\;\;\ \\ \label{T3} \sqrt{-g}J_{t}^{r}&\equiv&\frac{\tilde{T}
_{D7}r^5\dot{\phi}\phi^{\prime}}{\sqrt{1+r^2(\phi^{\prime})^2/6-L^4\dot{\phi}^2/6r^2}}
=\tilde{T}_{D7}r_0^4\omega^2.
\end{eqnarray}
Using (\ref{T1})--(\ref{T3}), we can derive the total energy and energy flux
of the D-brane system. The total energy of the D7-brane in the above configuration
is given by:
 
\begin{equation}
\label{Eden1}
E=\tilde{T}_{D7}\int_{r_0}^{\infty}{\frac{dr\,(r^{10}-r_0^8L^4\overline{\omega}^2)}
{r^2\sqrt{(r^8-r_0^8)(r^2-L^4\overline{\omega}^2)}}}.
\end{equation}
It is clear from (\ref{Eden1}) that when $r\rightarrow r_0$, the energy density of
the flavor brane, given by (\ref{T1}), becomes very large and blows up precisely at
the minimal extension, $r=r_0$, at the IR scale of conformal and chiral flavor symmetry
breaking. Thus we  conclude that at the IR scale of symmetries breakdown, $r=r_0$, the
backreaction of the D7-brane to the supergravity metric is non-negligibly large and
forms a black hole centered at the IR scale $r=r_0$ in the bulk. The black hole size
should grow as the energy is pumped into it from the D7-brane steadily. In order to
obtain this energy flux, we use the components of the energy--stress tensor 
(\ref{T1})--(\ref{T3}). We note that when the minimal radial extension is positive 
definite ($r_0>0$) and spin is turned on ($\omega>0$), the component (\ref{T3}) is
non-vanishing and hence we compute the time evolution of the total energy as:

\begin{equation}
\label{dE}
\dot{E}=\frac{d}{dt}\int{dr\sqrt{-g}J_{t}^{t}}=\int{dr\partial_r(\sqrt{-g}J_{t}^{r})}
=\sqrt{-g}J_{t}^{r}|_{r=r_0}^{\infty}=\tilde{T}_{D7}r_0^4\omega^2-\tilde{T}_{D7}r_0^4
\omega^2=0.
\end{equation}
Though the total energy is time-independent, relation (\ref{dE}) shows
that the energy (\ref{T3}) per unit time is injected at the boundary $r=\infty$
by some external system and the equal energy (\ref{T3}) is dissipated from the
IR into the bulk. Such dissipation from the D7-brane to the bulk will create a
black hole in the bulk. Thus, by this flow of energy form the probe to the bulk,
we conclude by duality that the energy from the flavor sector will eventually
dissipate into the gauge theory. In order to see explicitly this injection of
energy in the stationary solution, one may set UV and IR cut offs for the rotating
D7-brane solution and let the brane configuration extend from $r=r_{\text{IR}}=r_0$
to  $r=r_{\text{UV}}\gg r_{\text{IR}}$. We note from (\ref{T3}) that at $r_{\text{IR}}$
and $r=r_{\text{UV}}$ the value of $T_t^r$ is non-vanishing, which shows the 
presence of energy flux: The incoming energy flux from $r=r_{\text{UV}}$ is
given by  (\ref{T3}) and equals that of outgoing at $r=r_{\text{IR}}$, whereat
the energy does not get reflected back but its backreaction will form a black
hole intaking the injected energy.

It is also instructive to inspect the parameter dependence of the solution. 
Inspection of (\ref{T1}) shows that increasing the value of $\omega$, while
keeping the other parameters fixed, slightly shifts the minimum of the energy
density located near the IR scale of conformal and chiral flavor symmetry 
breaking; it also increases the scale of the density away from the IR scale,
as expected (see Fig.\,\ref{fig:KWp1}). Inspection of (\ref{T1}) also shows
that increasing the value of $r_0$, while keeping the other parameters fixed,
shifts the minimum of the energy density, but leaves its scale unchanged
(see Fig.\,\ref{fig:KWp1}). In any case, the energy density starts from its
infinite value at the IR scale, then decreases until its minimum value is 
reached, from where it increases to finitely small values away from the IR
scale. Thus we conclude that varying the parameters of the theory leaves the
overall behavior of the energy density unchanged, with the density blowing up
at the IR scale, where the backreaction is non-negligible, and remaining
finitely small away from the IR scale, where the backreaction can neglected.

\section{Rotating D7-branes in $adS_5\times T^{1,1}$ in the presence of world
volume gauge fields} 

In this section we aim to see how our analysis of the previous section gets modified
at finite baryon density. In the presence of $N_f$ flavors, the gauge theory posses
a global $U(N_f)\simeq SU(N_c)\times U(1)_q$ symmetry. The $U(1)_q$ counts the net
number of quarks, that is, the number of baryons times $N_c$. In the supergravity
description, this global symmetry corresponds to the $U(N_f)$ gauge symmetry on the
world volume of the $N_f$ D7-brane probes. The conserved currents associated with the
$U(N_f)$ symmetry of the gauge theory are dual to the gauge fields, $A_{\mu}$, on the
D7-branes. Hence the introduction of a chemical potential $\mu$ or a non-vanishing
$n_B$ for the baryon number the gauge theory corresponds to turning on the diagonal
$U(1)\subset U(N_f)$ gauge field, $A_{\mu}$ on the world volume of the D7-branes.
We may describe external electric and magnetic fields in the field theory, coupled to
anything having $U(1)$ charge, by introducing non-normalizable modes for $A_{\mu}$
in the supergravity theory (e.g. see ref.\,\cite{Kruczenski:2003be,O'Bannon:2008bz}).

\subsection{Induced metric and temperature in the presence of electric field}

In this subsection we will study D7-branes spinning with an angular frequency $\omega$,
and with a $U(1)$ world volume gauge field $A_{\mu}$. We note that in order to have the
gauge theory at finite chemical potential or baryon number density, it suffices to turn
on the time component of the gauge field, $A_t$. By symmetry considerations, one may take
$A_t=A_t(r)$. As we shall briefly discuss below, a potential of this form will support an
electric field.

We will consider an ansatz for the D7-brane world volume field, $\phi(r,t)=\omega t+f(r)$
and now $F_{ab}=F_{rt}=\partial_r A_t$. With this ansatz, it is straightforward to find the
components of the induced world volume metric on the D7-brane, $g_{ab}^{D7}$, and compute
the determinant in (\ref{AC}), giving the DBI Lagrangian as:

\begin{eqnarray}
\label{dbiac2}
S_{D7}=-\tilde{T}_{D7}\int{dr dt\,r^3\sqrt{1-\frac{L^4\dot{\phi}^2}{6r^2}+
\frac{r^2\,(\phi^{\prime})^2}{6}-(A^{\prime}_t(r))^2}}.
\end{eqnarray}
Here we note that by setting $\dot{\phi}=\omega=0$ and $A_{t}(r)=0$,
our action (\ref{dbiac2}) reduces to that of the Kuperstein--Sonnenschein
model, (\ref{KSA}). As in the previous section, following the
Kuperstein--Sonnenschein model reviewed in Sec.\,2, we set $\theta=\pi/2$
and restrict brane motion to the equator of the $S^2$ sphere. Thus, in our
set up we let, in addition, the probe rotate about the equator of the $S^2$,
as before, and further turn on a non-constant world volume electric gauge
field on the  probe. 

The equation of motion from the  action  (\ref{dbiac2}) then take the form:

\begin{eqnarray}
\label{D7eq}
\frac{\partial}{\partial r}\Bigg[\frac{r^5\,\phi^{\prime}}{\sqrt{1+
\frac{r^2\,(\phi^{\prime})^2}{6}-\frac{L^4\, \dot{\phi}^2}{6r^2}-
(A^{\prime}_t(r))^2}}\Bigg]&=&\frac{\partial}{\partial t}\Bigg[\frac{L^4
\,r\,\dot{\phi}}{\sqrt{1+\frac{r^2\,(\phi^{\prime})^2}{6}-\frac{L^4\,
\dot{\phi}^2}{6r^2}-(A^{\prime}_t(r))^2}}\Bigg],\;\;\ \\ \frac{\partial}
{\partial r}\Bigg[\frac{r^3\, A^{\prime}_t(r)}{\sqrt{1+\frac{r^2\,
(\phi^{\prime})^2}{6}-\frac{L^4\, \dot{\phi}^2}{6r^2}-(A^{\prime}_t(r))^2}}
\Bigg]&=& 0.
\end{eqnarray}
Taking the large radii limit, $r\rightarrow \infty$, gives the approximate
solution of the last equation as: $A_t(r)\simeq \mu-a_B/r^2$. Here $\mu$ is
the chemical potential and $a_B$ is the vacuum expectation value of baryon
number. We would like to remark that our solution for $A_t$ is of expected
form. We note that an electric field will be supported by a potential of the
from $A_t\simeq r^{-2}$. As this is a rank one massless field in $adS$, it 
must correspond to a dimension four operator or current in the gauge theory.
This is just what one would expect from an R--current, to which gauge fields
correspond. Consider rotating solutions of the form
\begin{eqnarray}
\label{rotsol2}
\phi(r,t)&=&\omega t+f(r),\;\;\;\;\ f(r)=\sqrt{6}r_0^4\int_{r0}^{r}{\frac{dr}{r}
\sqrt{\frac{1-L^4\,\overline{\omega}^2/r^2-(A^{\prime}_t(r))^2}{r^8-r_0^8}}}.
\end{eqnarray}
Here we note that by setting $\overline{\omega}=\omega/\sqrt{6}=0$
and $A_{t}(r)=0$, our solution (\ref{rotsol2}) reduces to that of
the Kuperstein--Sonnenschein model,  \cite{Kuperstein:2008cq}, 
reviewed in Sec.\,2 (see Eq.\,(\ref{KUPSol.})). It is also clear
that the above rotating solution has three free parameters, the
minimal radial extension $r_0$, the angular velocity $\omega$,
and the VEV of baryon number $a_B$. The solution (\ref{rotsol2})
describes brane motion with spin and non-constant world volume
electric gauge field turned on, with the brane starting and ending
up at the boundary. The brane comes down from the UV boundary at
infinity, bends at the minimal extension in the IR, and backs up
the boundary. We also note that when $r$ is large, the behavior of
the derivative of $f(r)$ with respect to $r$, denoted $f_r(r)$, with
and without $\omega$ and world volume electric field is the same
(see Fig.\,\ref{fig:KWgp2}).  This shows that in such limit the 
derivative of $f_r(r)$ integrates to the $\phi(r)$ of the 
Kuperstein--Sonnenschein model (see Sec.\,2) with the boundary
values $\phi_{\pm}$ in the asymptotic UV limit, $r\rightarrow\infty$
(see also Fig.\,\ref{fig:KWgp2}). However, we note that in the (opposite)
IR limit, i.e., when $r$ is small, the behavior of $f_r(r)$ does depend
on $\omega$, as before. We also note here that turning on, in addition,
the world volume electric field merely changes the scale but leaves the 
behavior of $f_r(r)$ in the IR unchanged. Inspection of (\ref{rotsol2})
shows that in the IR only for certain values of $\omega$ the behavior
of $f_r(r)$ with compares to that of without $\omega$ (see Fig.\,\ref{fig:KWgp2}).
This shows that in the IR and within specific range of $\omega>0$ the behavior of
$f_r(r)$ (here) compares to that of $\phi^{\prime}(r)$ in the Kuperstein--Sonnenschein
model (see Sec.\,2), where  $\phi^{\prime}(r)\rightarrow\infty$ in the IR limit
$r\rightarrow r_0$, consistent with U-like embedding.

Again, to derive the induced metric on the D7-brane in this configuration, we put
the solution (\ref{rotsol2}) into the background metric (\ref{10DKWmet}) and obtain:

\begin{eqnarray}
\label{ind2}
ds_{ind.}^2&=&-\frac{1}{3L^2}(3r^2-L^4\omega^2)dt^2\notag\\ && +\frac{L^2}{r^2}
\left[\frac{3r^2(r^8-r_0^8)+r_0^8(6r^2(1-(A^{\prime}_t(r))^2)-L^4\omega^2)}{3r^2
(r^8-r_0^8)}\right]dr^2\notag\\ &&+\frac{2L^2 \omega r_0^4}{3r^2}\sqrt{\frac{6r^2
(1-(A^{\prime}_t(r))^2)-L^4\,\omega^2}{r^8-r_0^8}}drdt\notag\\ &&+\frac{r^2}{L^2}
(dx^2+dy^2+dz^2)+\frac{L^2}{3}\left[\frac{1}{2}(\Omega_1^2+\Omega_2^2)+\frac{1}{3}
\Omega_3^2\right]\notag\\ &&-\frac{L^2}{3}\left[\omega\Omega_1 dt+ \frac{r_0^4}{r^2}
\sqrt{\frac{6r^2(1-(A^{\prime}_t(r))^2)-L^4\,\omega^2}{r^8-r_0^8}}\Omega_1 dr\right].
\end{eqnarray}
Here we note that by setting $\omega=0$ and $A_{t}(r)=0$, our induced
world volume metric (\ref{ind2}) reduces to that of the 
Kuperstein--Sonnenschein model, \cite{Kuperstein:2008cq}, reviewed in
Sec.\,2. In this case, for $r_0=0$ the induced world volume metric is
that of $adS_5\times S^3$ and the dual gauge theory describes the 
conformal and chiral symmetric phase. On contrary, for $r_0>0$ the 
induced world volume metric has no $adS$ factor and the conformal 
invariance of the dual gauge theory must be broken in such case, as
before. In order to find the world volume horizon and Hawking temperature,
we first eliminate the relevant cross term, as before. To eliminate the
relevant cross-term, we now consider a coordinate transformation:

\begin{equation}
\tau=t-\omega\,L^4 r_0^4 \int{\frac{dr\,(6r^2(1-(A^{\prime}_t(r))^2)-L^4
\,\omega^2)^{1/2}}{r^2(3r^2-L^4 \omega^2)(r^8-r_0^8)^{1/2}}}.
\end{equation}
The induced metric on the rotating D7-brane(s) then takes the form:

\begin{eqnarray}
\label{indA}
ds_{ind.}^2 &=&-\frac{(3r^2-L^4 \omega^2)}{3L^2}d\tau^2\notag\\ &&+
\frac{L^2}{r^2}\left[\frac{(3r^2-L^4 \omega^2)(r^8-r_0^8)+r_0^8(6r^2
(1-(A^{\prime}_t(r))^2)-L^4 \omega^2)}{(3r^2-L^4 \omega^2)(r^8-r_0^8)}
\right]dr^2\notag\\ && +\frac{r^2}{L^2}(dx^2+dy^2+dz^2)+\frac{L^2}{3}
\left[\frac{1}{2}(\Omega_1^2+\Omega_2^2)+\frac{1}{3}\Omega_3^2\right]
\notag\\ &&-\frac{L^2}{3}\left[\omega\Omega_1d\tau+\frac{3\,r_0^4}
{3r^2-L^4 \omega^2} \sqrt{\frac{6r^2(1-(A^{\prime}_t(r))^2)-L^4\,\omega^2}
{r^8-r_0^8}}\Omega_1 dr\right].
\end{eqnarray}
Here we note that for $r_0=0$ the induced world volume metric (\ref{indA}) has no
horizon, such that $-g_{tt}=g^{rr}=0$, and therefore not given by the black hole
geometry. On contrary, for $r_0>0$ the induced world volume horizon is described
by the horizon equation of the form:

\begin{figure}[t]
\begin{center}
\epsfig{file=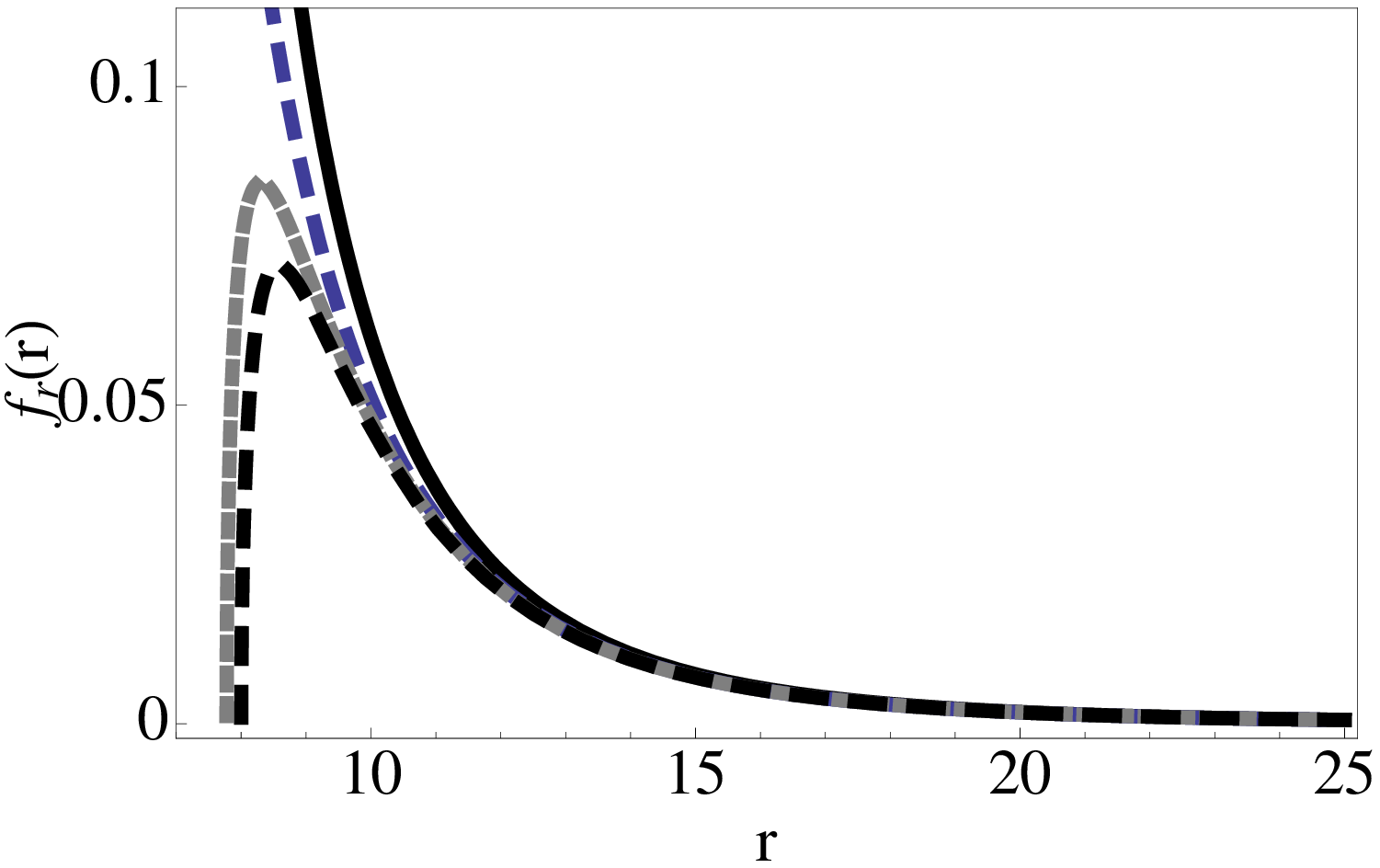,width=.50\textwidth}~~\nobreak
\epsfig{file=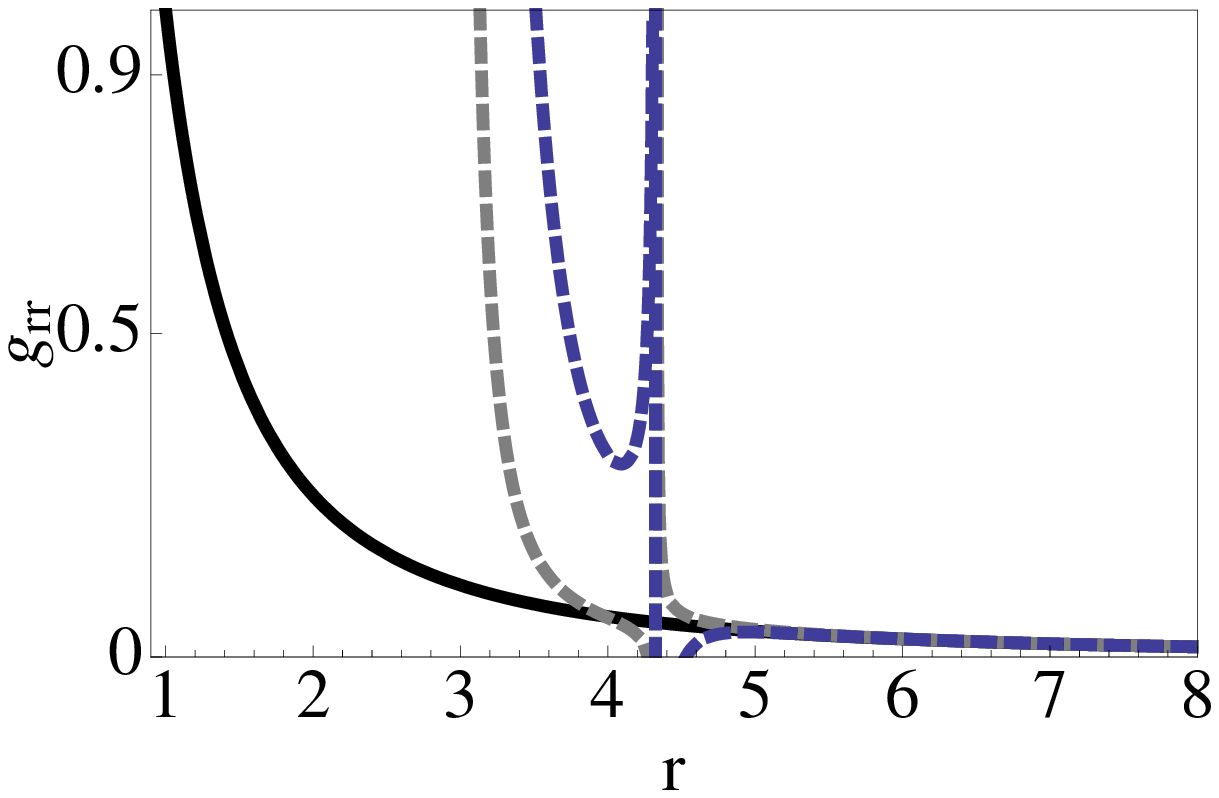,width=.50\textwidth}
\caption{[Left] The behavior of the derivative of the world
volume field with respect to $r$ with $L=1, r_0=7$, $\omega=0$
(black-solid), $\omega=5$ (blue-dashed), $\omega=7$ (gray-dashed),
and $\omega=8.5$ (black-dashed). [Right] The behavior of the 
$g_{rr}$ component of the induced world volume metric with 
$L=1,r_0=\omega=a_B=0$ (black-solid), $L=1, r_0=3, \omega=7.5, a_B=20$
(gray-dashed), and $L=1, r_0=3,\omega=7.5, a_B=60$ (blue-dashed).}
\label{fig:KWgp2}
\end{center}
\end{figure}

\begin{equation}
\label{HorEq2}
H(r)=r^2(r^8-r_0^8)(3r^2-L^4\omega^2)=0.
\end{equation}
This equation is identical with Eq.\,(\ref{HorEq1}) and therefore has the
same two real positive definite zeros (see also Fig.\,\ref{fig:KWgp2}). We
therefore note that though the world volume electric field modifies the
$g_{rr}$ component of the induced world volume metric on the rotating probe  
(cf. Fig.\,\ref{fig:KWgp2} $\&$ Fig.\,\ref{fig:KWgp}) the induced world
volume horizon described by (\ref{HorEq2}) remains unchanged compared to 
(\ref{HorEq1}). Therefore, the thermal horizon of the induced world volume
black hole geometry can be again identified with the solution of the 
horizon equation (\ref{HorEq2}) as $-g_{tt}=g^{rr}=H(r)=3r_h^2-L^4\omega^2=0$.
We therefore conclude at this point that when $r_0$ is positive definite
($r_0>0$) and spin is turned on ($\omega>0$), the induced world volume metric
on the rotating probe  admits a thermal horizon growing from the minimal 
extension $r_0\neq 0 $ with increasing the angular velocity, unaffected by
the presence of the world volume electric field.

\begin{figure}[t]
\begin{center}
\epsfig{file=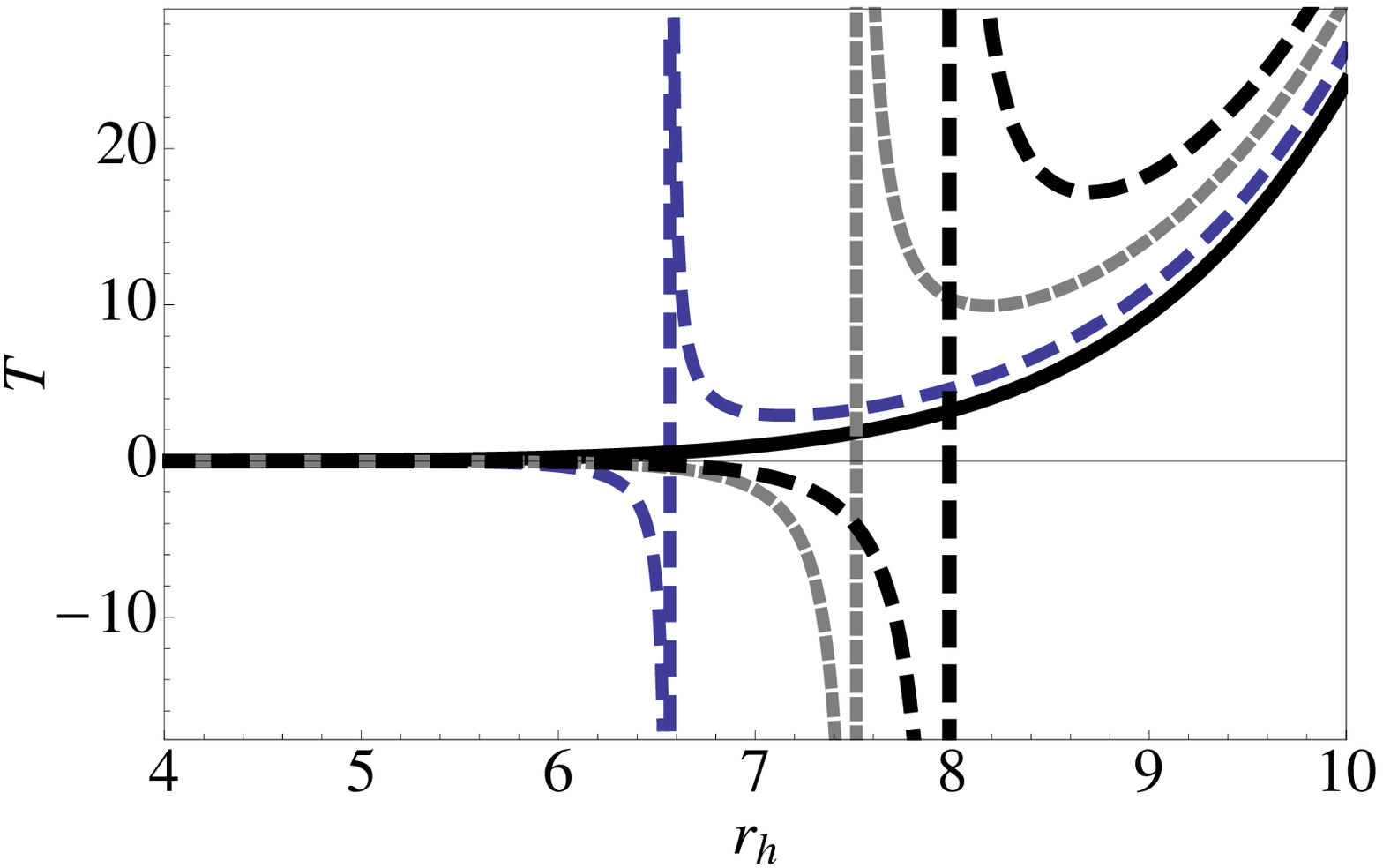,width=.5\textwidth}~~\nobreak
\epsfig{file=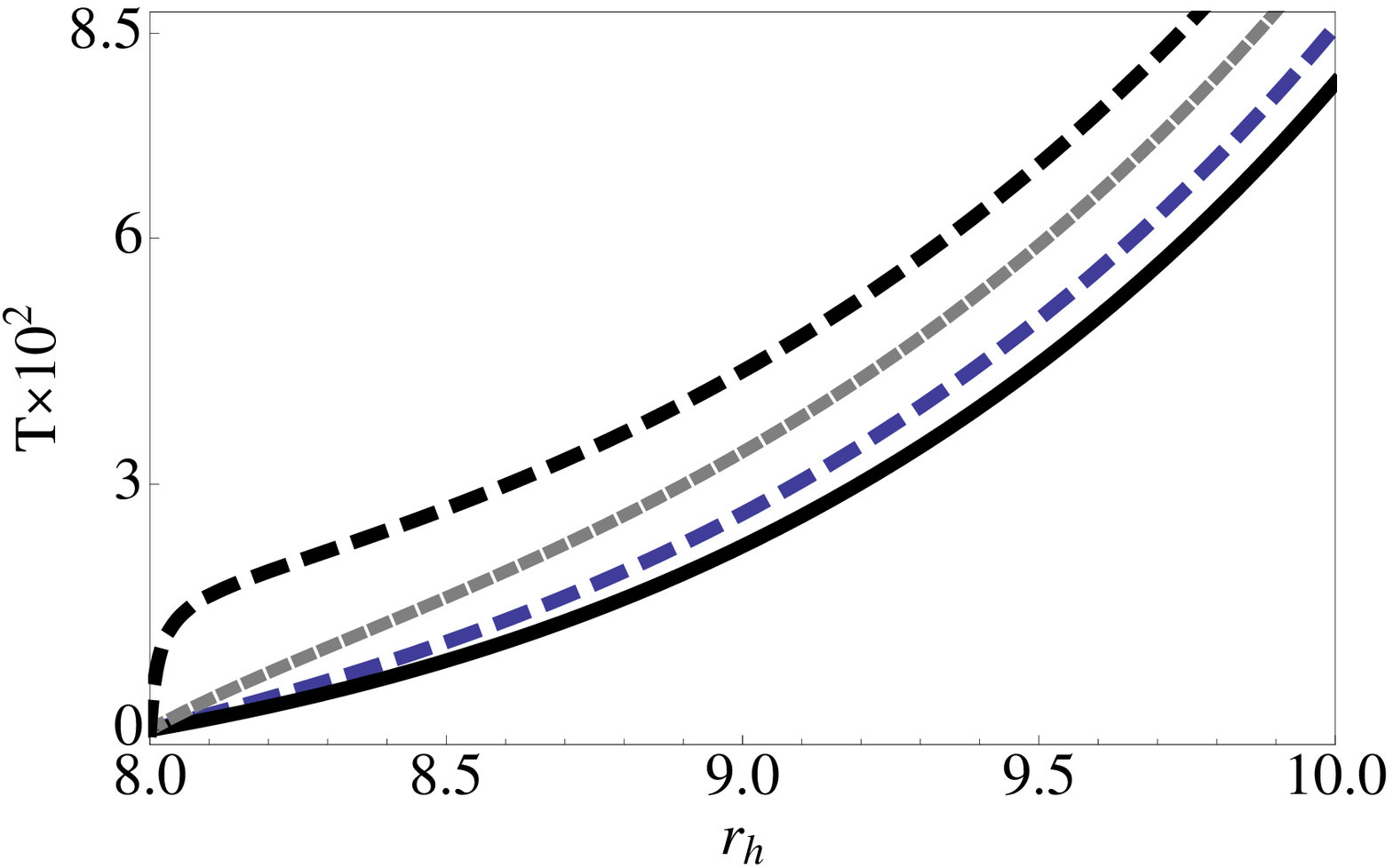,width=.5\textwidth}
\epsfig{file=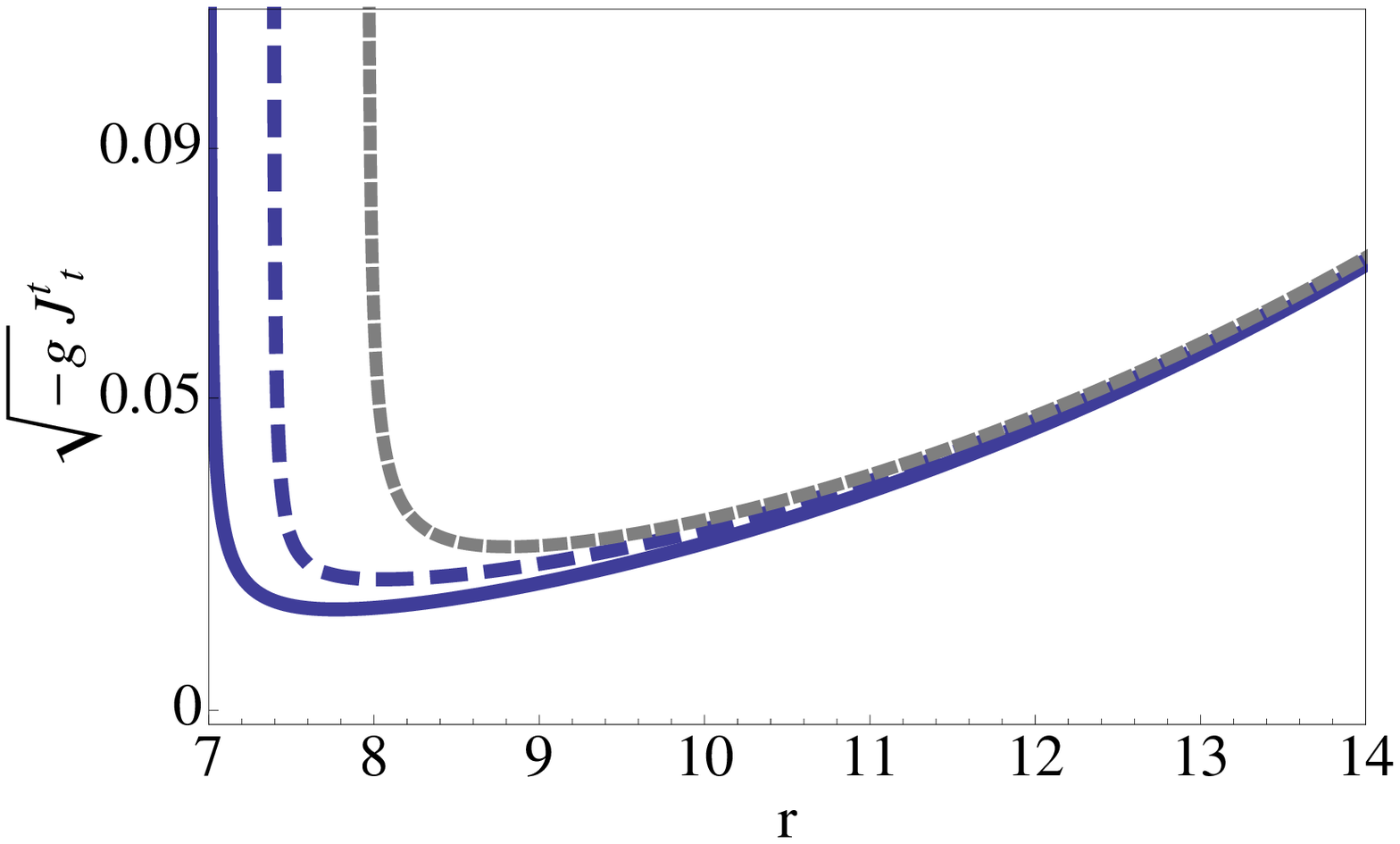,width=.5\textwidth}~~\nobreak
\epsfig{file=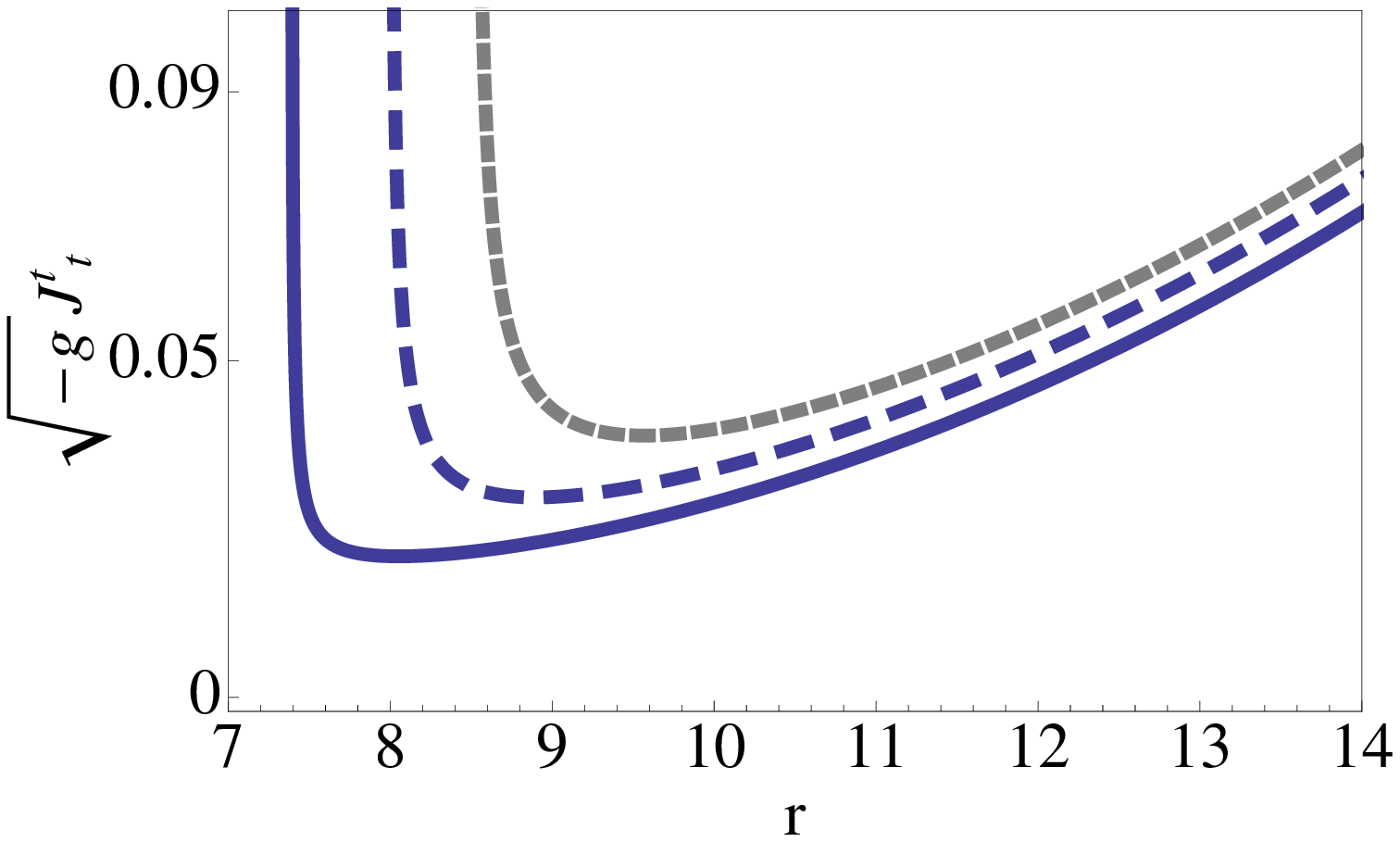,width=.5\textwidth}
\caption{[Up-Left] The behavior of the temperature
with $a_B=0$ (solid), $a_B=100$ (blue-dashed), $a_B=150$
(gray-dashed), $a_B=180$ (black-dashed), $L=10$ and 
$r_0=4$. [Up-Right] The behavior of the temperature with
$a_B=0$ (solid), $a_B=100$ (blue-dashed), $a_B=150$ 
(gray-dashed), $a_B=180$ (black-dashed), $L=10$ and $r_0=8$.
[Down-Left] The behavior of the energy-density, $\sqrt{-g}J_t^t$,
with $a_B=0$ (solid), $a_B=200$ (blue-dashed), $a_B=250$ (gray-dashed),
$r_0=7$, and $\omega=1$. [Down-Right] The behavior of the 
energy-density with $\omega=1$ (solid), $\omega=5$ (blue-dashed),
$\omega=6.5$ (gray-dashed), $r_0=7$ and  $a_B=200$. In these
cases, $L=\alpha^{\prime}=1$ and $g_s=0.1$.}
\label{fig:KWgp3}
\end{center}
\end{figure}

The Hawking temperature can be found from this induced metric in the form:

\begin{equation}
\label{TH2}
T=\frac{(g^{rr})^{\prime}}{4\pi}\bigg|_{r=r_h}=\frac{3r_h^3(r_h^8-r_0^8)}
{2\pi L^2 r_0^8[6r_h^2(1-(A^{\prime}_t(r))^2)-L^4\omega^2]}=\frac{r_h^7
(r_h^8-r_0^8)}{2\pi L^2 r_0^8(r_h^6-8a_B^2)}.
\end{equation}
From (\ref{TH2}) it seems that the temperature of the black hole solution increases
with growing horizon size, as it should. It also seems that increasing the vacuum
expectation value of the baryon number density $a_B$ increases the temperature whereas
at $a_B=0$ we obtain the temperature (\ref{TH1}), as we should. However, careful 
inspection of (\ref{TH2}) shows that the temperature of the solution has three distinct
branches including two obvious classes of black hole solutions.  First, for $a_B\neq 0$
and the rest of parameters fixed, the temperature $T$ decreases with increasing horizon
size $r_h$. Here we note that as the horizon starts to grow, the temperature $T$ becomes
immediately negative\footnote{Here we note that in other interesting study in the
literature, ref.\,\cite{Nakamura:2013yqa}, using similar D-brane systems, it has been
shown that when just an electric field is turned on (in place of rotation, or R--charge, 
considered here), in certain codimensions, the induced world volume temperature of the
probe is given by a decreasing function of the electric field (see Eq.\,(24) in 
ref.\,\cite{Nakamura:2013yqa})}. Then, $T$ peeks off very sharply, producing a divergent
type behavior, where at some point it hits zero, before growing into positive values.
Finally, $T$ decreases positively until it reaches a non-zero minimum 
(see Fig.\,\ref{fig:KWgp3}). In this case, the temperature of the solution seems to go
more or less with the inverse of the horizon. Thus in the vicinity of the zero point the
temperature of the solution decreases with  increasing horizon size. These `small' black
holes have the familiar behavior of five-dimensional black holes in asymptotically flat
spacetime, since their temperature decreases with increasing horizon size. Second, for 
$a_B\neq 0$ and the rest of parameters fixed, the other class of black hole solution appears,
as the horizon size continues to grow from its value that settles the positive minimum of $T$.
In this case, the temperature of the solution only increases with increasing horizon size, 
similar to the case with $a_B=0$. In this regime, we also note that increasing the VEV of $a_B$
increases the scale of the temperature, relative to the case $a_B=0$ (see Fig.\,\ref{fig:KWgp3}).
We also note that when we set $r_0<1$, the scale of $T$  increases, but its behavior remains 
unchanged (as in Fig.\,\ref{fig:KWgp3}).

However, the above behavior of the temperature changes when the value of
$r_0$ is set larger, while the rest of parameters are fixed as before. 
Inspection of  (\ref{TH2}) shows that when $r_0$ is fixed by larger values,
the temperature $T$ increases continuously with increasing the horizon size
$r_h$, and that increasing the VEV of $a_B$ increases the temperature $T$ 
(see Fig.\,\ref{fig:KWgp3}). We note though that at sufficiently large horizon
radii in this case  $T$ scales much less than in the previous where $r_0$ was
chosen smaller (see Fig.\,\ref{fig:KWgp3}). We therefore conclude that when
the external world-volume electric field is turned on, $a_B\neq 0$, the scale
and behavior of the temperature changes according to the size of the modulus
$r_0$. Namely, for small values of $r_0$, the theory admits two classes of 
temperatures. There is one branch where the temperature increases with increasing
horizon size and there is another where the temperature decreases with increasing
horizon size. At relatively larger values of $r_0$, there is only one branch where
the temperature increases continuously with growing horizon size. In both cases, at
sufficiently large horizon size the scale of the temperature increases with increasing
the VEV of the baryon number density $a_B$.

We also note that by considering the backreaction of this solution to the KW SUGRA
background, one naturally expects the D7-brane to form a very small black hole in KW,
describing a locally thermal gauge field theory in the probe limit, as before.
Accordingly, the rotating D7-brane describes a thermal object in the dual gauge field
theory. In the KW example here, the system is dual to  $\mathcal{N}=1$ gauge theory
coupled to a quark in the presence of an external electric field. Since the gauge
theory itself is at zero temperature while the quark is at finite temperature $T$,
(\ref{TH2}), the system is in  non-equilibrium steady state. However, as
discussed below, the energy from the flavor sector will eventually dissipate into
the gauge theory, as before.

In the above analysis, the backreaction of the D7-brane to the supergravity background
has been neglected since we considered the probe limit. It is instructive to see to what
extend this can be justified. The components of the stress--energy tensor of the D7-brane
take the form\footnote{See footnote 6.}:

\begin{eqnarray}
\label{TA1}
\sqrt{-g}J_{t}^{t}&\equiv&\frac{\tilde{T}_{D7}\,r^3(1+r^2(\phi^{\prime})^2/6)}
{\sqrt{1+r^2(\phi^{\prime})^2/6-L^4\dot{\phi}^2/6r^2-(A^{\prime})^2}}\notag\\
&=&\frac{\tilde{T}_{D7}(r^4(r^{10}-L^4\overline{\omega}^2r_0^8)-4r_0^8a_B^2)}
{r^4\sqrt{(r^8-r_0^8)(r^6-L^4\overline{\omega}^2 r^4-4a_B^2)}},\;\;\;\;\;\;\;\;\
\\ \label{TA2}\sqrt{-g}J_{r}^{r}&\equiv&-\frac{\tilde{T}_{D7}\,
r^3(1-L^4\dot{\phi}^2/6r^2)}{\sqrt{1+r^2(\phi^{\prime})^2/6-L^4\dot{\phi}^2/6r^2
-(A^{\prime})^2}}\notag\\ &=&-\tilde{T}_{D7}(r^2-L^4\overline{\omega}^2)
\sqrt{\frac{r^8-r_0^8}{r^6-L^4\overline{\omega}^2 r^4-4a_B^2}}, \\  \label{TA3}
\sqrt{-g}J_{t}^{r}&\equiv&\frac{\tilde{T}_{D7}r^5\dot{\phi}\phi^{\prime}}
{\sqrt{1+r^2(\phi^{\prime})^2/6-L^4\dot{\phi}^2/6r^2-(A^{\prime})^2}}=
\tilde{T}_{D7}r_0^4\omega.
\end{eqnarray}
Using (\ref{TA1})--(\ref{TA3}), we can derive the total energy and energy
flux of the D-brane system. The total energy of the D7-brane in the above
configuration is given by:

\begin{equation}
\label{Eden2}
E=\tilde{T}_{D7}\int_{r_0}^{\infty}{\frac{dr\,(r^4(r^{10}-L^4\overline{\omega}^2
r_0^8)-4r_0^8a_B^2)} {r^4\sqrt{(r^8-r_0^8)(r^6-L^4\overline{\omega}^2 r^4-4a_B^2)}}}.
\end{equation}
It is straightforward to see that when the electric field is tuned off,
(\ref{TA1})--(\ref{Eden2}) reduce to (\ref{T1})--(\ref{Eden1}), as they should.
It is also clear from (\ref{Eden2}) that when $r\rightarrow r_0$, the energy
density of the flavor brane, given by (\ref{TA1}), becomes very large and
blows up precisely at the minimal extension, $r=r_0$, at the IR scale of 
conformal and chiral flavor symmetry breaking, independent from the presence
of the world volume electric field. Thus we  conclude that at the IR scale of
symmetries breakdown, $r=r_0$, the backreaction of the D7-brane to the 
supergravity metric is non-negligibly large and forms a black hole centered
at the IR scale $r=r_0$ in the bulk, as before. The black hole size should grow
as the energy is pumped into it from the D7-brane steadily. In order to obtain
this energy flux, we use the components of the energy--stress tensor 
(\ref{TA1})--(\ref{TA3}). We note that when the minimal radial extension is 
positive definite ($r_0>0$) and spin is turned on ($\omega>0$), the component
(\ref{TA3}) is non-vanishing and hence we compute the time evolution of the total
energy as:
\begin{equation}
\label{dEA}
\dot{E}=\frac{d}{dt}\int{dr\sqrt{-g}J_{t}^{t}}=\int{dr\partial_r(\sqrt{-g}J_{t}^{r})}
=\sqrt{-g}J_{t}^{r}|_{r=r_0}^{\infty}=\tilde{T}_{D7}r_0^4\omega^2-\tilde{T}_{D7}r_0^4
\omega^2=0.
\end{equation}
Here we note that when the electric field is turned on, (\ref{TA3}) and energy dissipation
(\ref{dEA}) remain unchanged, compared with (\ref{T3}) and (\ref{dE}). Therefore,
independent from the electric field, the energy dissipation from the brane into
the bulk will form a black hole, centered at the IR scale $r=r_0$, as before (see Sec.\,3.1).
Thus, by this flow of energy form the probe to the bulk, we conclude by duality
that the energy from the flavor sector will eventually dissipate into the gauge
theory, independent from the electric field. To see this external injection of energy in our
stationary rotating solution, we may again adopt UV/IR cut offs and find from (\ref{TA3}) 
that, independent from the electric field, at $r=r_{\text{IR}}$ the energy is unreflected
back but its backreaction will form a black hole, absorbing the injected energy, as before
(see Sec.\,3.1).

It is also instructive to inspect the parameter dependence of the theory
in the presence of the electric field. Inspection of (\ref{TA1}) shows 
that when the electric field in turned on, the scale and behavior of the
energy density remains unchanged (see Fig.\,\ref{fig:KWgp3}). Inspection of
(\ref{TA1}) shows that increasing $a_B$ by relatively large values, while
keeping the rest of parameters fixed, merely shifts the minimum of the energy
density (see Fig.\,\ref{fig:KWgp3}). Inspection of (\ref{TA1}) also shows that
increasing the value of the angular velocity $\omega$, while keeping the rest
of parameters fixed, increases the energy density, as expected
(see Fig.\,\ref{fig:KWgp3}). However, the energy density is increased such that
the backreaction remains negligibly small away from the blowing up point 
(see Fig.\,\ref{fig:KWgp3}). Thus we conclude that in the presence of the electric
field  varying the parameters of the theory leaves the behavior of the energy
density more or less unchanged, with the density blowing up at the IR scale, where
the backreaction is non-negligible, and  remaining finitely small elsewhere, where
the backreaction is negligible.

\subsection{Induced metric and temperature in the presence of magnetic field}

In this subsection we will study D7-branes spinning with an angular
frequency $\omega$, and with a $U(1)$ world volume gauge field $A_{\mu}$,
but in place of turning on an electric field, we will be interested
in a magnetic field, which we may introduce through adding to our 
D7-brane ansatz the constant magnetic field $F_{xy}=B$. In the gauge
theory, one can identify $F_{xy}$ as a constant $U(1)$ magnetic field
pointing in the z direction. The relevance of introducing $B$ is that
at zero temperature, zero mass, and zero R--charge chemical potential,
gauge/gravity calculations have shown that such a $U(1)$ $B$-field
triggers spontaneous breaking of the R--symmetry \cite{O'Bannon:2008bz}.
However, we are interested in solutions at finite R--charge chemical
potential and $U(1)$ magnetic field. Such solutions describe states in
the gauge theory with degenerate hypermultiplet fields, a finite R--charge
density, and spontaneous breaking of the R--symmetry.

We will consider an ansatz for the D7-brane world volume field
$\phi(r,t)=\omega t+f(r)$ and now $F_{xy}=B$. With this ansatz,
it is straightforward to find the components of the induced world
volume metric on the D7-brane, $g_{ab}^{D7}$, and compute the 
determinant in (\ref{AC}), giving the DBI Lagrangian as:
\begin{eqnarray}
\label{dbiac3}
S_{D7}=-\tilde{T}_{D7}\int{dr dt\, r^3\sqrt{\bigg(1+\frac{L^4\hat{B}^2}{r^4}
\bigg)\bigg(1-\frac{L^4\dot{\phi}^2}
{6r^2}+\frac{r^2\,(\phi^{\prime})^2}{6}\bigg)}}.
\end{eqnarray}
Here we note that by setting $\dot{\phi}=\omega=0$ and $\hat{B}=(2\pi
\alpha^{\prime})B=0$, our action (\ref{dbiac3}) reduces to that of the
Kuperstein--Sonnenschein model, (\ref{KSA}). As in the previous sections,
following the Kuperstein--Sonnenschein model reviewed in Sec.\,2, we set
$\theta=\pi/2$ and restrict brane motion to the equator of the $S^2$ sphere.
Thus, in our set up we let, in addition, the probe rotate about the equator
of the $S^2$, as before, and further turn on a constant world volume magnetic
gauge field on the probe. 

The equation of motion from the action (\ref{dbiac3}) takes then the form:

\begin{eqnarray}
\label{D7eq}
\frac{\partial}{\partial r}\left[\frac{r^5\,(1+L^4\hat{B}^2/r^4)\,
\phi^{\prime}}{\sqrt{(1+L^4\hat{B}^2/r^4)(1-L^4\dot{\phi}^2/6r^2
+r^2(\phi^{\prime})^2/6)}}\right]&=&\notag \\ \frac{\partial}{\partial t}
\left[\frac{L^4 r\,(1+L^4\hat{B}^2/r^4)\dot{\phi}}{\sqrt{(1+L^4\hat{B}^2/r^4)
(1-L^4\dot{\phi}^2/6r^2+r^2\,(\phi^{\prime})^2/6}}\right].
\end{eqnarray}

Consider rotating solutions of the form
\begin{eqnarray}
\label{rotsol3}
\phi(r,t)&=&\omega t+f(r),\;\;\;\;\ f(r)=\sqrt{6}r_0^4\int_{r_0}^{r}{\frac{dr}{r}
\sqrt{\frac{1-L^4\overline{\omega}^2/r^2}{r^6(r^2+L^4 \hat{B}^2)-r_0^8}}}.
\end{eqnarray}
Here we note that by setting $\overline{\omega}=\omega/\sqrt{6}=0$
and $\hat{B}=0$, our solution (\ref{rotsol3}) reduces to that of
the Kuperstein--Sonnenschein model,  \cite{Kuperstein:2008cq}, 
reviewed in Sec.\,2 (see Eq.\,(\ref{KUPSol.})). It is also clear
that the above rotating solution has three free parameters, the
minimal radial extension $r_0$, the angular velocity $\omega$,
and the value of the magnetic field $B$. The solution (\ref{rotsol3})
describes brane motion with spin and constant world volume
magnetic gauge field turned on, with the brane starting and ending
up at the boundary. The brane comes down from the UV boundary at
infinity, bends at the minimal extension in the IR, and backs up
the boundary. We also note that when $r$ is large, the behavior of
the derivative of $f(r)$ with respect to $r$, denoted $f_r(r)$, with
and without $\omega$ and world volume electric field is the same
(see Fig.\,\ref{fig:KWgp3}).  This shows that in such limit the 
derivative of $f_r(r)$ integrates to the $\phi(r)$ of the 
Kuperstein--Sonnenschein model (see Sec.\,2) with the boundary
values $\phi_{\pm}$ in the asymptotic UV limit, $r\rightarrow\infty$
(see also Fig.\,\ref{fig:KWgp3}). However, we note that in the (opposite)
IR limit, i.e., when $r$ is small, the behavior of $f_r(r)$ does depend
on $\omega$ and $B$. Inspection of (\ref{rotsol3}) shows that in the IR
only for certain values of $\omega$ and $B$ the behavior of $f_r(r)$ with
compares to that of without $\omega$ and $B$ (see Fig.\,\ref{fig:KWgp3}).
This shows that in the IR and within specific range of $\omega,B>0$ the
behavior of $f_r(r)$ (here) compares to that of $\phi^{\prime}(r)$ in the
Kuperstein--Sonnenschein model (see Sec.\,2), where  
$\phi^{\prime}(r)\rightarrow\infty$ in the IR limit $r\rightarrow r_0$,
consistent with U-like embedding.

As before, to derive the induced metric on the D7-brane in this configuration,
we put the rotating solution (\ref{rotsol3}) into the background metric 
(\ref{10DKWmet}) and obtain:

\begin{eqnarray}
\label{ind3}
ds_{ind.}^2&=&-\frac{1}{3L^2}(3r^2-L^4\omega^2)dt^2\notag\\ &&
+\frac{L^2}{r^2}\left[\frac{3r^2(r^6(r^2+L^4\hat{B}^2)-r_0^8)+r_0^8
(6r^2-L^4\omega^2)}{3r^2(r^6(r^2+L^4\hat{B}^2)-r_0^8)}\right]dr^2
\notag\\ &&+\frac{2L^2 \omega\, r_0^4}{3r^2}\sqrt{\frac{6r^2-L^4\,
\omega^2}{r^6(r^2+L^4\hat{B}^2)-r_0^8}}drdt+\frac{r^2}{L^2}(dx^2
+dy^2+dz^2)\notag\\ &&+\frac{L^2}{3}\left[\frac{1}{2}(\Omega_1^2+
\Omega_2^2)+\frac{1}{3}\Omega_3^2- \omega\Omega_1 dt- \frac{r_0^4}
{r^2}\sqrt{\frac{6r^2-L^4\,\omega^2}{r^6(r^2+L^4\hat{B}^2)-r_0^8}}
\Omega_1 dr\right].\notag\\
\end{eqnarray}
Here we note that by setting $\omega=0$ and $B=0$, our induced world volume
metric (\ref{ind3}) reduces to that of the Kuperstein--Sonnenschein
model, \cite{Kuperstein:2008cq}, reviewed in Sec.\,2. In this case,
for $r_0=0$ the induced world volume metric is that of $adS_5\times S^3$
and the dual gauge theory describes the conformal and chiral symmetric
phase. On contrary, for $r_0>0$ the induced world volume metric has no
$adS$ factor and the conformal invariance of the dual gauge theory must
be broken in such case, as before. In order to find the world volume
horizon and Hawking temperature, we first eliminate the relevant cross
term, as before. To eliminate the relevant cross-term, we now consider
a coordinate transformation:

\begin{equation}
\tau=t-\omega\,L^4 r_0^4 \int{\frac{dr\,(6r^2-L^4\,\omega^2)^{1/2}}
{r^2(3r^2-L^4 \omega^2)(r^6(r^2+L^4\hat{B}^2)-r_0^8)^{1/2}}}.
\end{equation}
The induced metric on the rotating D7-brane(s) then takes the form:

\begin{eqnarray}
\label{indB}
ds_{ind.}^2 &=&-\frac{(3r^2-L^4 \omega^2)}{3L^2}d\tau^2\notag\\
&&+\frac{L^2}{r^2}\left[\frac{(3r^2-L^4 \omega^2)
(r^6(r^2+L^4\hat{B}^2)-r_0^8)+r_0^8(6r^2-L^4 \omega^2)}
{(3r^2-L^4 \omega^2)(r^6(r^2+L^4\hat{B}^2)-r_0^8)}\right]dr^2
\notag\\ && +\frac{r^2}{L^2}(dx^2+dy^2+dz^2)+\frac{L^2}{3}
\left[\frac{1}{2}(\Omega_1^2+\Omega_2^2)+\frac{1}{3}\Omega_3^2
\right]\notag\\ &&-\frac{L^2}{3}\left[\omega\Omega_1
d\tau+\frac{3\,r_0^4}{3r^2-L^4 \omega^2} 
\sqrt{\frac{6r^2-L^4\,\omega^2}{r^6(r^2+L^4\hat{B}^2)-r_0^8}}
\Omega_1 dr\right].
\end{eqnarray}
Here we note that for $r_0=0$ the induced world volume metric (\ref{indA}) has no
horizon, such that $-g_{tt}=g^{rr}=0$, and therefore not given by the black hole
geometry. On contrary, for $r_0>0$ the induced world volume horizon is described
by the the horizon equation of the form:

\begin{figure}[t]
\begin{center}
\epsfig{file=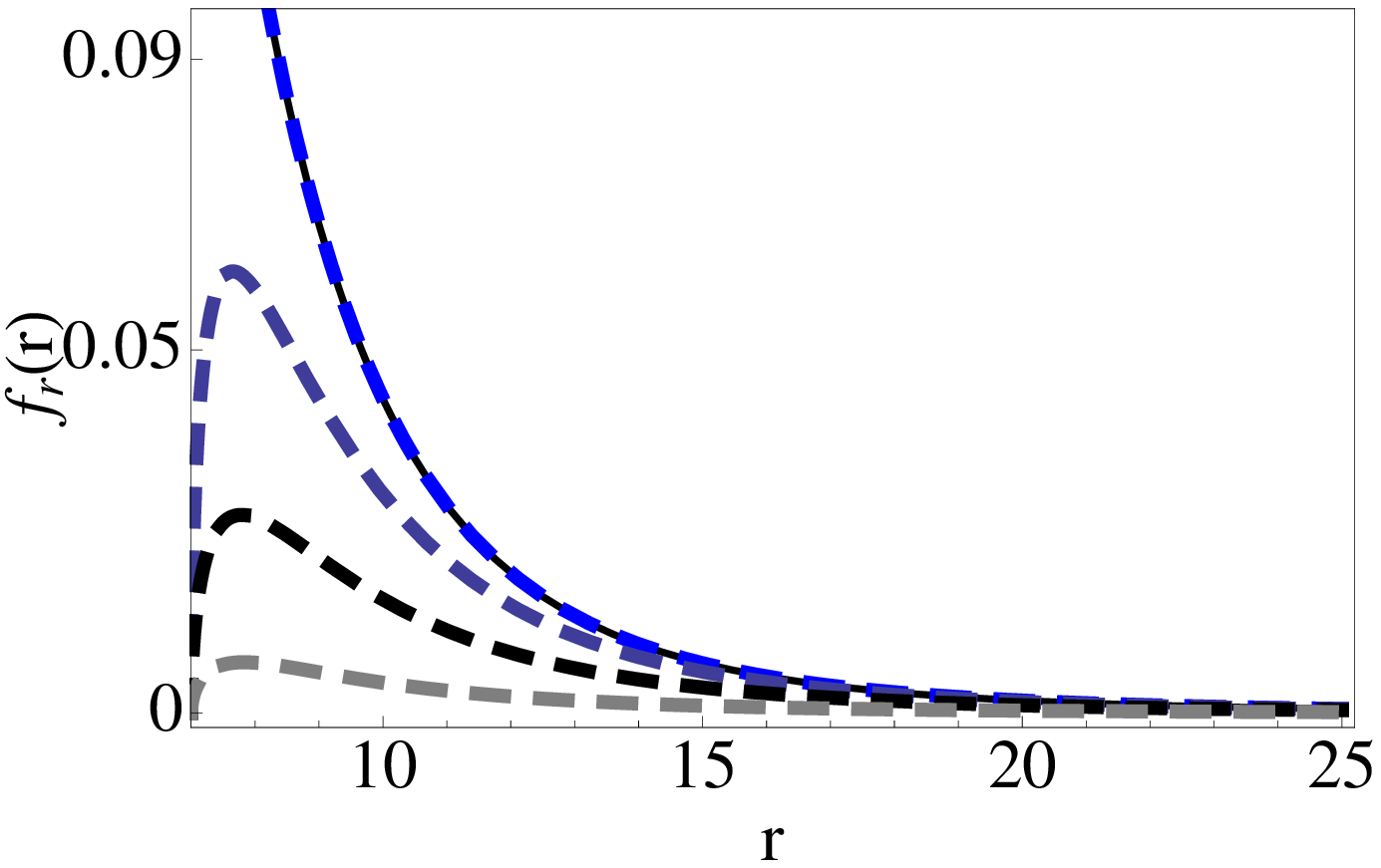,width=.50\textwidth}~~\nobreak
\epsfig{file=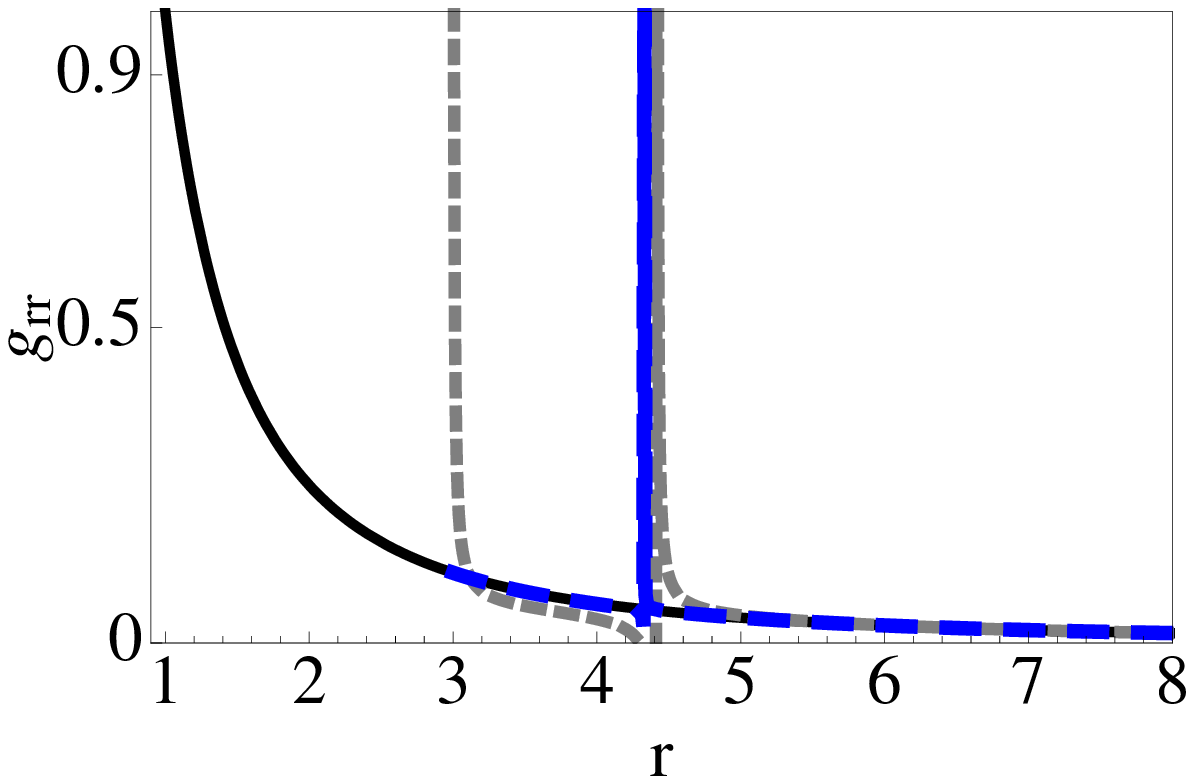,width=.50\textwidth}
\caption{[Left] The behavior of the derivative of the world volume
field with respect to $r$ with $L=1, r_0=7, \omega=7,$, $B=0$ 
(black-solid), $B=0.05$ (blue-dashed), $B=10$ (dark blue-dashed),
$B=25$ (black-dashed), and $B=100$ (gray-dashed). [Right] The 
behavior of the $g_{rr}$ component of the induced world volume 
metric with $L=1,r_0=\omega=B=0$ (black-solid), $L=1, r_0=3,
\omega=7.5$, $B=0.05$ (gray-dashed), and $B=25$ (blue-dashed).}
\label{fig:KWgp40}
\end{center}
\end{figure}

\begin{equation}
\label{HorEq3}
H(r)=r^2(r^6(r^2+L^4B^2)-r_0^8)(3r^2-L^4\omega^2)=0.
\end{equation}
This equation, unlike (\ref{HorEq2}), is not the same as (\ref{HorEq1}),
though it has some similar features. We note that the world volume
magnetic field modifies the $g_{rr}$ component of the induced world
volume metric (see Fig.\,\ref{fig:KWgp40}) as well as the induced 
world volume horizon described by (\ref{HorEq3}), though the thermal
horizon can be identified with the solution of (\ref{HorEq3}) as
$-g_{tt}=g^{rr}=H(r)=3r^2-L^4\omega^2=0$, as before. However, we note
that by setting $B$ large, the induced world volume horizon described
by (\ref{HorEq3}) appears to start from zero. In other words, by setting
$B$ large and demanding the world volume horizon to start from the
minimal radial extension $r_0$, the horizon equation (\ref{HorEq3})
implies $r_0=0$. This may indicate that when the world volume magnetic
field is turned on, the world volume horizon has to start from the
conifold point, $r=r_0=0$, whereby the induced world volume horizon is
inside the $adS$ horizon. But, we saw that for $r_0=0$ the induced world
volume metric (\ref{indB}), has no  thermal horizon, to begin with. And,
we note that in U-like embeddings the point $r_0=0$ is outside the validity
range of the induced world volume metric (\ref{indB}). Nevertheless, closer 
inspection of the horizon equation (\ref{HorEq3}) shows that when $B$ is set
small, the world volume horizon starts nearly from $r_0\neq0$. We therefore
conclude at this point that when the minimal extension is positive
definite ($r_0>0$) and spin is turned on ($\omega>0$) and the world volume
magnetic field ($B=const.$) is induced, the induced world volume metric
admits a thermal horizon growing almost from $r_0$ with increasing the angular
velocity, provided that the magnetic field is weak.

The Hawking temperature can be found from this induced metric in the form:

\begin{equation}
\label{TH3}
T=\frac{(g^{rr})^{\prime}}{4\pi}\bigg|_{r=r_h}=\frac{3r_h^3[r_h^4
(r_h^4+L^4\hat{B}^2)-r_0^8]}{2\pi L^2 r_0^8(6r_h^2-L^4\omega^2)}
=\frac{r_h[r_h^4(r_h^4+L^4\hat{B}^2)-r_0^8]}{2\pi L^2 r_0^8}.
\end{equation}
From (\ref{TH3}) it is clear that the temperature of the black hole solution
increases with growing horizon size. It is also clear that increasing the 
magnetic field $B$ increases the temperature whereas at $B=0$ we obtain the
temperature (\ref{TH1}), as we should. We also note that in the presence of
the magnetic field, $B\neq0$, at $r_h=r_0$ there is a minimum temperature, 
below which there is no world volume black hole solution on the probe. This
is unlike in previous examples where the minimum temperature was zero. The
minimum temperature, in this case, goes with quadric of $B$ and with inverse
cube of $r_0$, $T_0=(L\,\hat{B})^2/2\pi r_0^3$. This shows that for fixed 
values of $L$ the size of the minimum temperature $T_0$ increases with 
increasing $B$ and decreasing $r_0$. Inspection of (\ref{TH3}) also shows
that when the magnetic field is turned on, the temperature of the world volume
black hole solution increases continuously with growing horizon size 
(see Fig.\,\ref{fig:KWgp4}). This qualitative behavior is independent from the
choice of parameters.  Comparison with Sec.\,1 also shows that increasing the
value of the magnetic field, while $r_0$ is fixed, affects the scale of the 
temperature by more or less the same amount as decreasing $r_0$ does, while the
magnetic field is turned off, with the same large hierarchy
$T_{r_0>1}/T_{r_0<1}\simeq 10^8$ (see Fig.\,\ref{fig:KWgp4} and 
Fig.\,\ref{fig:KWp1}). Comparison with Sec.\,1 also shows that that in the presence
of the magnetic field the behavior of the temperature remains the same. This is in
contrast with Sec.\,3 where the presence of the electric field modified the behavior
of the temperature  (see Fig.\,\ref{fig:KWgp4}, Fig.\,\ref{fig:KWgp3} and 
Fig.\,\ref{fig:KWp1}). We therefore conclude that when the world-volume  magnetic
field $B$ is turned on,  the temperature $T$ grows from its minimum $T_0$, with the
scale of $T$ increased, while its behavior remains unchanged, as $B$ is increased at
any fixed $r_0$. 

We also note that by considering the backreaction of this solution to the KW SUGRA
background, one naturally expects the D7-brane to form a very small black hole in KW,
describing a locally thermal gauge field theory in the probe limit, as before.
Accordingly, the rotating D7-brane describes a thermal object in the dual gauge field
theory. In the KW example here, the system is dual to $\mathcal{N}=1$ gauge theory
coupled to a quark in the presence of an external magnetic field. Since the gauge
theory itself is at zero temperature while the quark is at finite temperature $T$,
given by (\ref{TH3}), the system is in  non-equilibrium steady state. However, as
discussed below, the energy from the flavor sector will eventually dissipate to the
gauge theory, as before.

In the above analysis, the backreaction of the D7-brane to the supergravity background
has been neglected since we considered the probe limit. It is instructive to see
to what extend this can be justified. The components of the stress--energy tensor of
the D7-brane take the form\footnote{See footnote 6.}:

\begin{eqnarray}
\label{TB1}
\sqrt{-g}J_{t}^{t}&\equiv&\frac{\tilde{T}_{D7}\,r^3(1+L^4\tilde{B}^2/r^4)^{1/2}
(1+r^2(\phi^{\prime})^2/6)}{\sqrt{1+r^2(\phi^{\prime})^2/6-L^4\dot{\phi}^2/6r^2}}
\notag\\&=&\frac{\tilde{T}_{D7}[r^6(r^4+L^4\tilde{B}^2)-r_0^8L^4\overline{\omega}^2]}
{r^4}\sqrt{\frac{r^4+L^4\tilde{B}^2}{[r^4(r^4+L^4\tilde{B}^2)-r_0^8][r^2-L^4
\overline{\omega}^2]}},\;\;\;\ \\ \label{TB2} \sqrt{-g}J_{r}^{r}&\equiv& 
-\frac{\tilde{T}_{D7}\,r^3(1+L^4\tilde{B}^2/r^4)^{1/2}(1-L^4\dot{\phi}^2/6r^2)}
{\sqrt{1+r^2(\phi^{\prime})^2/6-L^4\dot{\phi}^2/6r^2}}\notag\\ &=&-\frac{\tilde{T}_{D7}}
{r^2}\sqrt{(r^2-L^4\overline{\omega}^2)[r^4(r^4+L^4\tilde{B}^2)-r_0^8]},\\ \label{TB3}
\sqrt{-g}J_{t}^{r}&\equiv& \frac{\tilde{T}_{D7}r^5(1+L^4\tilde{B}^2/r^4)^{1/2}
\dot{\phi}\phi^{\prime}}{\sqrt{1+r^2(\phi^{\prime})^2/6-L^4\dot{\phi}^2/6r^2}} 
=\tilde{T}_{D7}r_0^4\omega.
\end{eqnarray}
Using (\ref{TB1})--(\ref{TB3}), we can derive the total energy and energy
dissipation of the system. The total energy of the D7-brane in the above
configuration is given by:

\begin{figure}[t]
\begin{center}
\epsfig{file=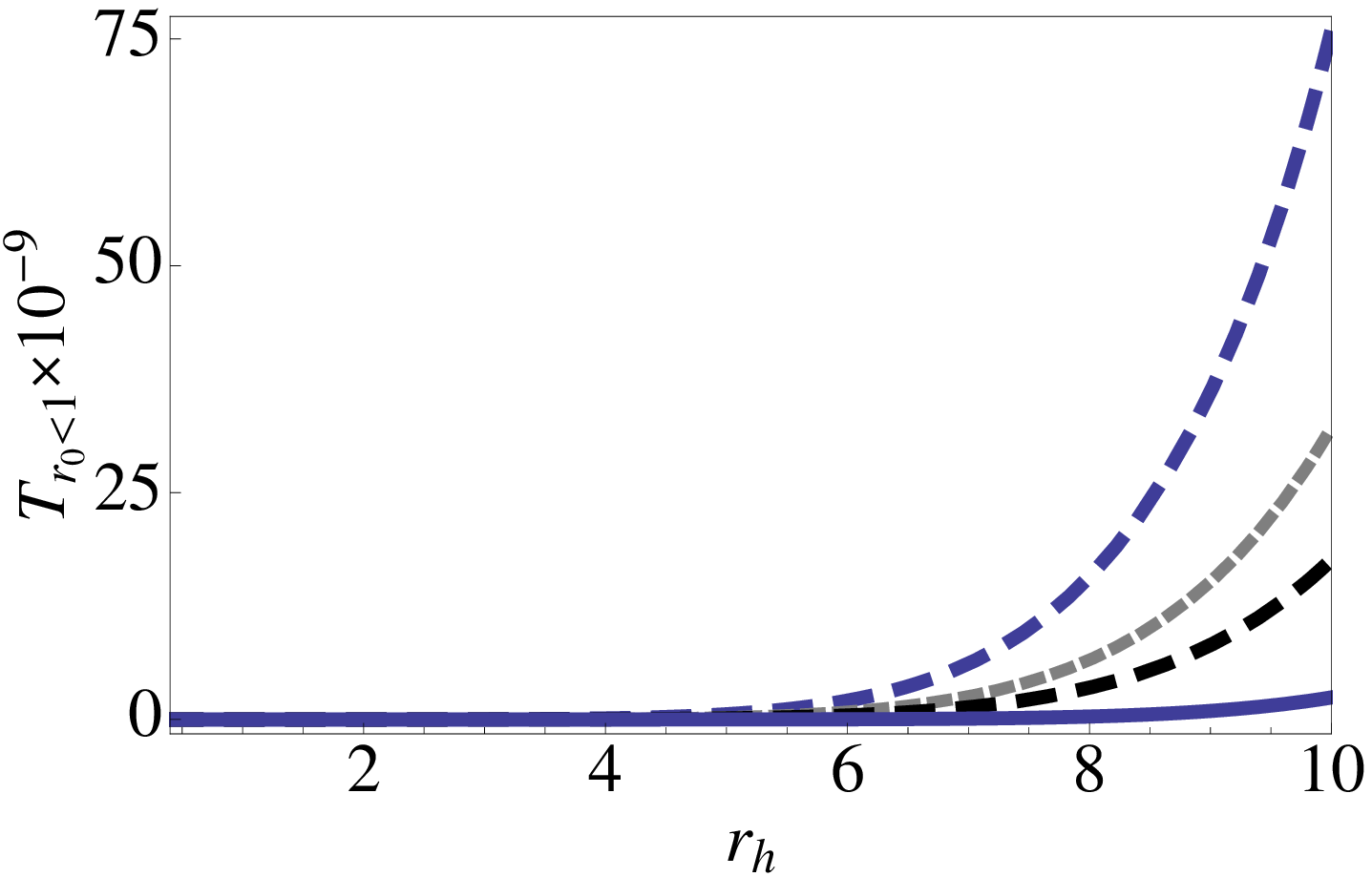,width=.5\textwidth}~~\nobreak
\epsfig{file=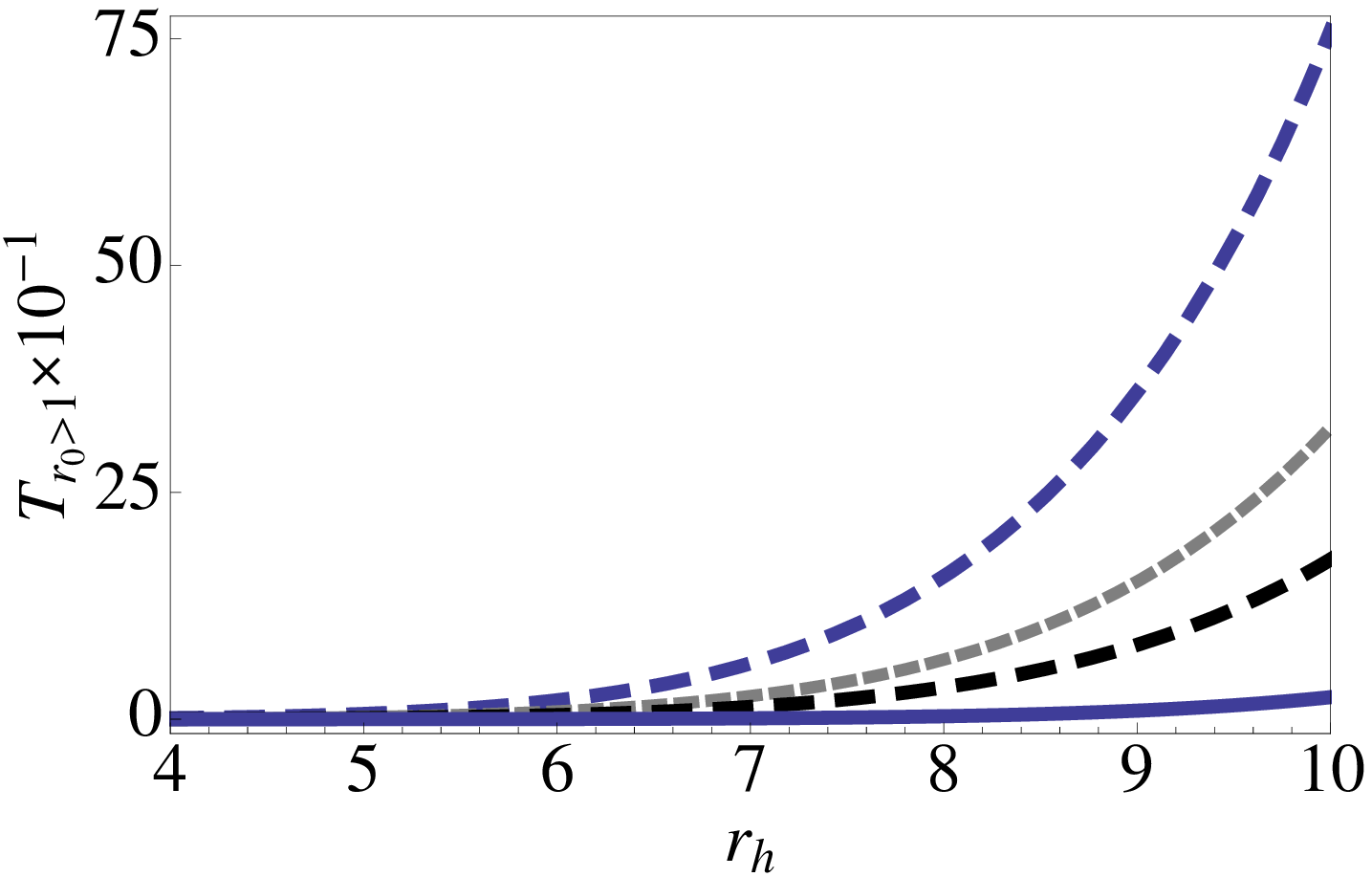,width=.5\textwidth}
\epsfig{file=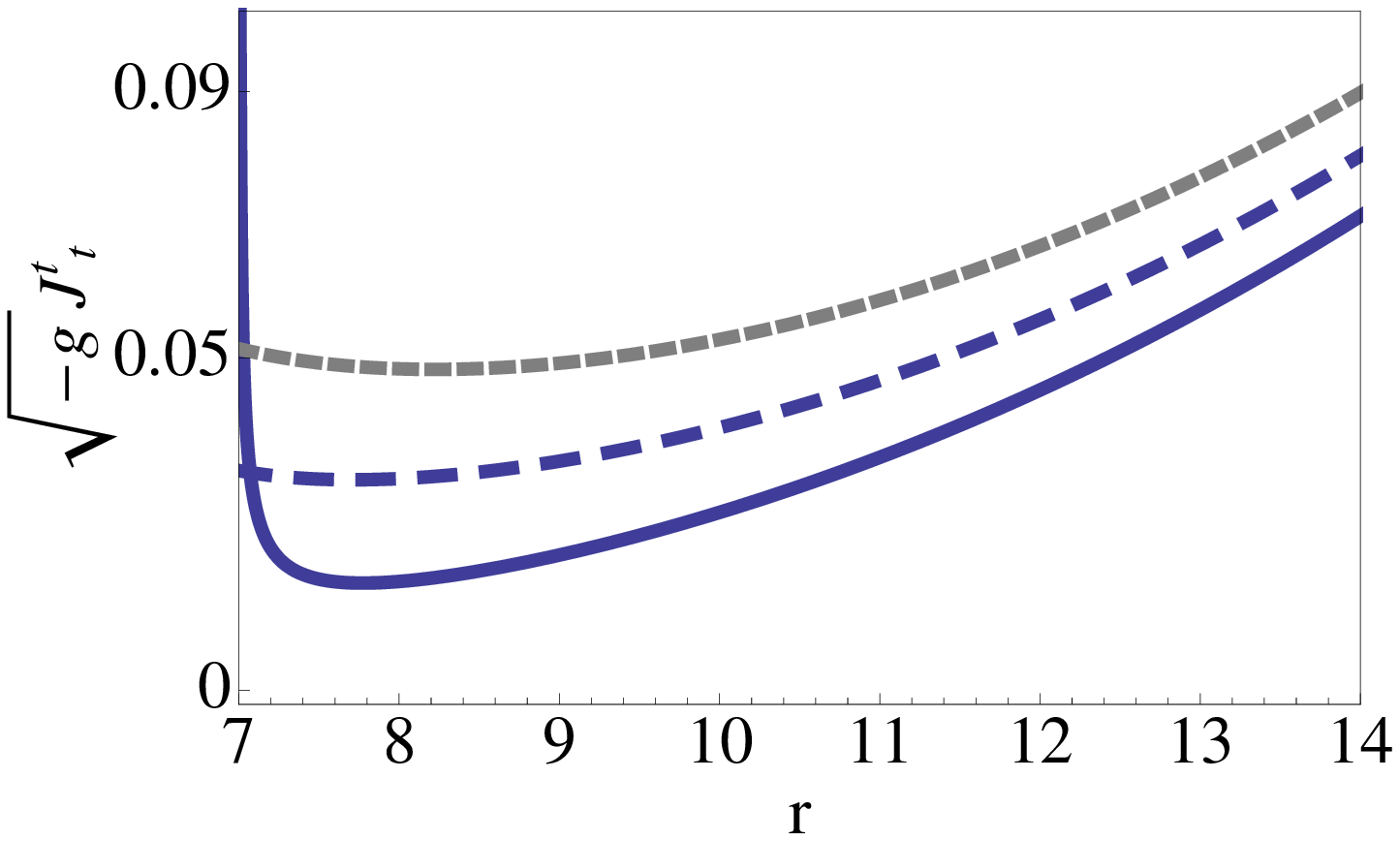,width=.5\textwidth}~~\nobreak
\epsfig{file=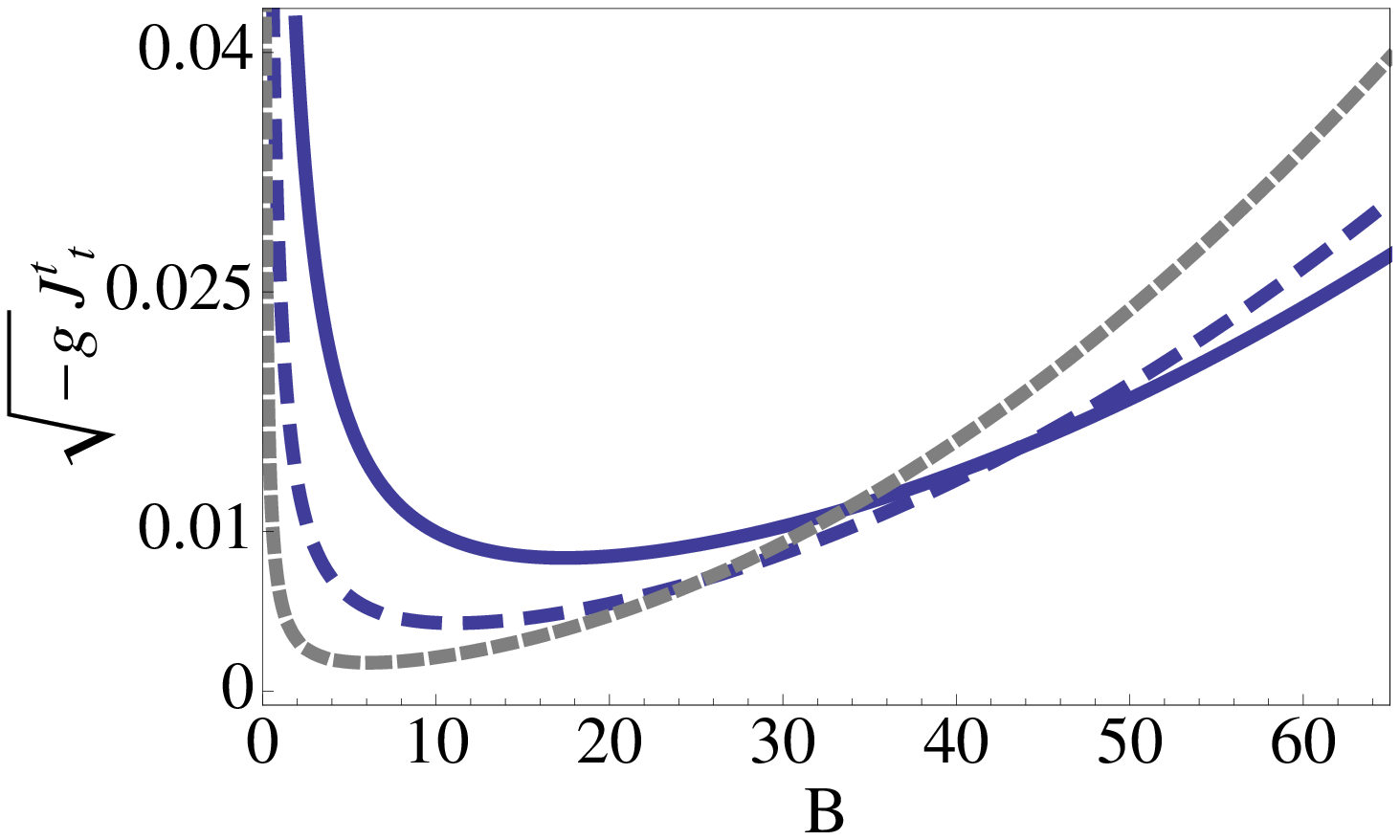,width=.5\textwidth}
\caption{[Up-Left] The behavior of the temperature with $B=0$ (solid),
$B=0.25$ (black-dashed), $B=0.35$ (gray-dashed), $B=0.55$ (blue-dashed),
$L=10$ and $r_0=4$. [Up-Right] The behavior of the temperature with 
$r_0=0.4$ and the same values of $B$. [Down-Left] The behavior of the 
energy-density, $\sqrt{-g}J_t^t$, with $B=0$ (solid), $B=70$ (dashed),
$B=100$ (gray-dashed), and $r_0=7$. [Down-Right] The behavior of the
energy-density with $r_0=5$ (solid), $r_0=4$ (dashed), $r_0=3$ 
(gray-dashed). Here: $L=\alpha^{\prime}=\omega=1$, and $g_s=0.1$.}
\label{fig:KWgp4}
\end{center}
\end{figure}

\begin{equation}
\label{Eden3}
E=\tilde{T}_{D7}\int_{r_0}^{\infty}{\frac{dr\,[r^6(r^4+L^4\tilde{B}^2)-r_0^8L^4
\overline{\omega}^2]}{r^4}\sqrt{\frac{r^4+L^4\tilde{B}^2}{[r^4(r^4+L^4\tilde{B}^2)
-r_0^8][r^2-L^4\overline{\omega}^2]}}}.
\end{equation}
It is easy to see that when the magnetic field is tuned off, $B=0$, (\ref{TB1})--(\ref{Eden3})
reduce to (\ref{T1})--(\ref{Eden1}), as they should. It is also clear form (\ref{Eden3}) that
in the presence of the magnetic field, $B \neq0$, the value of the energy density, given by
(\ref{TB1}), remains finite at the IR scale $r=r_0$, on contrary with the previous examples
(Secs.\,3.1 $\&$ 4.1). Nonetheless, closer inspection of the denominator of (\ref{TB1}) shows
that when the magnetic field $B$ is not too large, compared with other scales of the theory,
the energy density does blow up instead in the vicinity of the minimal extension $r=r_0$, nearby
the IR scale of conformal and chiral flavor symmetry breaking. Thus we conclude that close to the
IR scale of symmetries breakdown $r=r_0$, the backreaction of the D7-brane to the supergravity
background metric is non-negligibly large and forms a black hole centered nearby the IR scale
$r=r_0$ in the bulk. The black hole size should grow as the energy is pumped into it from the
D7-brane steadily. In order to obtain this energy flux, we note that the components of the 
energy--stress tensor are given by (\ref{TB1}--(\ref{TB3}). We note that when the minimal radial
extension is positive definite ($r_0>0$) and spin is turned on ($\omega>0$), the component
(\ref{TB3}) is non-vanishing and hence we compute the time evolution of the total energy as:

\begin{equation}
\label{dEB}
\dot{E}=\frac{d}{dt}\int{dr\sqrt{-g}J_{t}^{t}}=\int{dr\partial_r(\sqrt{-g}J_{t}^{r})}
=\sqrt{-g}J_{t}^{r}|_{r=r_0=0}^{\infty}=\tilde{T}_{D7}r_0^4\omega^2-\tilde{T}_{D7}r_0^4
\omega^2=0.
\end{equation}
Here we note that when the magnetic field is turned on, the component (\ref{TB3})
and energy dissipation relation (\ref{dEB}) remain unchanged, compared with (\ref{T3})
and (\ref{dE}), as in the previous example (Sec.\,3.1). Therefore, independent from
the magnetic field, the energy dissipation from the brane into the bulk will form
a black hole centered nearby the IR scale $r=r_0$, as before (see Secs.\,3.1 $\&$ 4.1).
Thus, by this flow of energy from the brane into the bulk, we conclude by duality that
the energy from the flavor sector will eventually dissipate into the gauge theory,
independent from the magnetic field. To see this external injection of energy in our
stationary rotating solution, we may again consider UV and IR cut offs and find from
(\ref{TB3}) that, independent from the magnetic field, at $r=r_{\text{IR}}$ the energy
is unreflected back but its backreaction will form a black hole, absorbing the injected
energy, as before (see Secs.\,3.1 $\&$ 4.1).

It is also instructive to inspect the parameter dependence of the theory
in the presence of the magnetic field. Inspection of (\ref{TB1}) shows that
when the magnetic field is turned on, the energy density increases 
(see Fig.\,\ref{fig:KWgp4}). However, inspection of (\ref{TB1}) shows that
increasing the magnetic field $B$ by relatively large values, while keeping
the rest of parameters fixed, increases the energy density and hence the 
backreaction, though such that the  backreaction remains negligibly small
away from the blowing up point  (see Fig.\,\ref{fig:KWgp4}). Inspection of
(\ref{TB1}) also shows that at the IR scale $r=r_0$ the energy densities 
with different values of $r_0$, and other parameters fixed, become degenerate
for certain values of the magnetic field (see Fig.\,\ref{fig:KWgp4}). Thus we
conclude that in the presence of the magnetic field varying the parameters of
the theory leaves the overall behavior of the energy density more or less
unchanged, with the density blowing up nearby the IR scale, where the 
backreaction is non-negligible, and remaining finitely small elsewhere, where
the backreaction is negligible.

\section{Discussion}

In this paper, we studied, in detail, the induced world volume metrics and Hawking
temperatures of rotating probe D7-branes in the Kuperstein--Sonnenschein holographic
model. We also studied, in detail, the energy--stress tensors and energy flow of 
rotating probe D7-branes in the Kuperstein--Sonnenschein model. By gauge/gravity
duality, the Hawking  temperatures on the rotating probe D7-branes in supergravity
correspond to the temperatures of flavored quarks in the gauge theory, and the energy
flow from the probe D7-brane into the system, to the energy dissipation from the flavor
sector into the gauge theory. Such non-equilibrium systems and their energy dissipation
have already been studied in the literature in conformal $adS$ setups. The aim of
this work was to extend such previous analyses to more general and realistic holographic
setups. The motivation of this work was to construct novel examples of non-equilibrium
steady states in holographic models where the conformal and chiral flavor symmetry of
the dual gauge theory get spontaneously broken.

We derived the induced world volume metrics on rotating probe D7-branes, with and without
the presence of worldvolume gauge fields, in the Kuperstein--Sonnenschein model. We showed
that when the minimal extension of the probe is positive definite and spin is turned on,
the induced world volume metrics on the rotating probe admit thermal horizons and Hawking
temperatures despite the absence of black holes in the bulk. We found that the scale
and behavior of the probe temperature are determined strongly by the size of the minimal
extension of the probe, and by the strength and sort of world volume gauge fields. By 
duality, we thus found the scale and  behavior of temperature in flavored quarks are 
strongly determined by the IR scale of symmetries breakdown, and by the strength and sort
of external fields. We noted that by considering the backreaction of such solutions to the
holographic background, one naturally expects the D7-brane to form a very small black hole
in the background, corresponding to a locally thermal gauge field theory in the probe limit.
We thus found the rotating probe D7-brane describing a thermal object in the dual gauge field
theory, including a quark, with or without the presence of external fields. Because the gauge
theory itself was at zero temperature while the quark was at finite temperature, we found that
such systems are in non-equilibrium steady states. However, we found that in the IR the 
backreaction is large and the energy will dissipate from the probe D7-brane into the bulk. By
duality, we thus found the energy dissipation from the flavor sector into the gauge theory.

In the absence of the world volume gauge fields, we found that the world volume
horizon starts from the minimal extension and grows with increasing the angular
velocity.  We found the temperature increasing steadily from zero with the horizon
size growing from the minimal extension. We also found that the scale of the
temperature increases/decreases dramatically, while its behavior remains unchanged,
when the minimal extension is decreased/increased. We thus found, in this example,
that the temperature is determined by the shape of flavor brane embedding configuration:
`sharp' $\emph{U}$-like  configurations--approaching $\emph{V}$-like--($r_0<1$)
admit higher temperatures than `smooth' $\emph{U}$-likes ($r_0>1$). This may be
expected, since by decreasing the minimal extension the conformal and chiral symmetric
phase is approached. However, we saw that when the configuration is $\emph{V}$-like
($r_0=0$), corresponding to the conformal and chiral symmetric phase, the induced world
volume metric on the rotating probe admits no thermal horizon and hence no Hawking
temperature. We noted that by considering the backreaction of this solution to the KW
SUGRA background, one naturally expects the D7-brane to form a very small black hole 
in KW, corresponding to a locally thermal gauge field theory in the probe limit.
Accordingly, we found the rotating probe D7-brane describing a thermal object in the
dual gauge field theory. In this example, the system was dual to  $\mathcal{N}=1$ gauge
theory coupled to a quark. Since the gauge theory itself was at zero temperature while
the quark is at finite temperature, we found, in this example, that the system is in 
non-equilibrium steady state. However, we then showed that the energy from the flavor
sector will eventually dissipate into the gauge theory. We first found from the 
energy--stress tensor that, independent from the choice of parameters, at the minimal
extension, at the IR scale of conformal and chiral flavor symmetry breakdown, the energy
density blows up and hence showed the backreaction in the IR is non-negligible. We then
showed from the energy--stress tensor that when the minimal extension is positive definite
and spin is turned on, the energy flux is non-vanishing and found the energy can flow from
the D7-brane into the bulk, forming, with the large backreaction, a black hole in the bulk. 
We also argued how this external injection of energy may be understood in our stationary
solutions. We considered UV and IR cut offs in our rotating  D7-brane system and noted
from the energy--stress tensor that the incoming energy from the UV equals the outgoing
energy from the IR where the large backreaction forms a black hole intaking the injected
energy. By gauge/gravity duality, we thus found, in this example, the energy dissipation
from the flavor sector into the gauge theory.

In the presence of the world volume electric field, we first found that both the behavior
as well as the scale of the temperature get modified. We found that turning on the
world volume electric field renders the temperature described by two distinct branches.
We found that there is one branch where the temperature increases and another where it
decreases with increasing the horizon size. While the former branch is rather of usual
form, the latter describes `small' black holes. However, we then varied the parameters of 
the theory and found that the two distinct branches appear only for certain set of 
parameters. We found that for relatively larger values of the minimal extension the 
temperature only increases with increasing horizon size. We also found that at 
sufficiently large horizon size, for any fixed value of the minimal extension, increasing
the VEV of the baryon density number increases the scale of the temperature. We noted
that by considering the backreaction of this solution to the KW SUGRA background, one
naturally expects the D7-brane to form a very small black hole in KW, corresponding to
a locally thermal gauge field theory in the probe limit. Accordingly, we found the 
rotating D7-brane describing a thermal object in the dual gauge field theory. In this
example, the system was dual to  $\mathcal{N}=1$ gauge theory coupled to a quark in the
presence of an external electric field. Since the gauge theory itself was at zero 
temperature while the quark is at finite temperature, we found, in this example, that the
system is in non-equilibrium steady state. However, we then showed that the energy from
the flavor sector will eventually dissipate into the gauge theory. We first found from
the  energy--stress tensor that, independent from the VEV of baryon density number and
the choice of other parameters, at the minimal extension, at the IR scale of conformal and
chiral flavor symmetry breakdown, the energy density blows up and so showed the backreaction
in the IR is non-negligible. We then showed from the energy--stress tensor that when the
minimal extension is positive definite and spin is turned on, the energy flux is 
non-vanishing and found the energy can flow from the brane into the bulk, forming, with
the large backreaction, a black hole in the bulk, independent from the presence of the
electric field. We also argued how this external injection of energy may be understood
in our stationary solutions, by setting UV and IR cut offs in our rotating D7-brane
system, as before. By gauge/gravity duality, we thus found, in this example, the energy
dissipation from the flavor sector into the gauge theory, independent from the electric field.

In the presence of the world volume magnetic field, we found that only the scale of
the temperature changes. We found that when the magnetic field is turned on, the
temperature has a minimum below which there is no black hole solution. We found the
temperature increasing steadily from its minimum with growing horizon size. We also
showed that increasing the magnetic field increases the scale of the temperature, but
leaves its behavior unchanged. We found the increase in the temperature due to
increasing the magnetic field similar to the increase in the temperature due to 
decreasing the minimal extension with the world volume gauge fields turned off. We
noted that by considering the backreaction of this solution to the KW SUGRA background,
one naturally expects the D7-brane to form a very small black hole in KW, corresponding
to a locally thermal gauge theory in the probe limit. Therefore, we found the rotating 
D7-brane describing a thermal object in the dual gauge field theory. In this example,
the system was dual to  $\mathcal{N}=1$ gauge theory coupled to a quark in the
presence of an external magnetic field. Since the gauge theory itself was at zero 
temperature while the quark is at finite temperature, we found, in this example, that the
system is in  non-equilibrium steady state. However, we then showed that the energy from
the flavor sector will eventually dissipate into the gauge theory. We first found from
the  energy--stress tensor that, for small values of the magnetic field and independent
from other parameters, near the minimal extension, near the IR scale of conformal
and chiral flavor symmetry breakdown, the energy density blows up and so showed the
backreaction in the IR is non-negligible. We then showed from the energy--stress tensor
that when the minimal extension is positive definite and spin is turned on, the energy 
flux is non-vanishing and found the energy can flow from the brane into the bulk, forming,
with the large backreaction, a black hole in the bulk, independent from the presence of
the magnetic field. We also argued how this external injection of energy may be understood
in our stationary solutions, by introducing UV and IR cut offs in our rotating D7-brane
system, as before. By gauge/gravity duality, we thus found, in this example, the energy
dissipation from the flavor sector into the gauge theory, independent from the magnetic
field.

We conclude that the induced world volume metrics on rotating probe D7-branes
in the Kuperstein--Sonnenschein model admit thermal horizons and Hawking 
temperatures in spite of the absence of black holes in the bulk. We conclude,
in particular, that this world volume black hole formation is controlled by a
single parameter, by the  minimal extension of the probe, which sets the IR scale
of conformal and chiral flavor symmetry breakdown. We conclude that the scale
and behavior of the temperature on the rotating probe D7-brane are determined,
in particular, strongly by the size of the minimal  extension of the brane, and
by the strength and sort of world volume brane gauge fields. By gauge/gravity 
duality, we thus conclude that the scale and behavior of the temperature in 
flavored quarks are determined, in particular, strongly by the IR scale of 
conformal and chiral flavor symmetry breaking, and by the strength and sort of
external fields. Since the gauge theory is at zero temperature and the flavored
quarks are at finite temperature, we conclude that such systems describe 
non-equilibrium steady states in the gauge theory of conformal and chiral flavor
symmetry breakdown. We conclude, however, that, independent from the 
presence of world volume gauge fields, the energy flows from the brane into the
system, forming, with the large backreaction in the IR, a black hole in the system.
By gauge/gravity duality, we thus conclude that, independent from the external
fields, the energy will eventually dissipate from the flavor sector into the gauge
theory. 

There are limitations to our work. In the examples we constructed, we
were unable to fully solve the brane equations of motion to determine
the explicit analytic form of the world volume brane fields, when conserved
angular motion and world volume gauge fields were turned on. Therefore, in our
examples, we were not able to provide full details about the probe brane solution
itself, when the angular velocity and world volume gauge fields were set non-vanishing.
Nevertheless, in order to derive the induced world volume metrics and compute the
induced world volume Hawking temperatures, we did not need to find the explicit
form of the world volume brane field. However, in all examples we constructed, we
deliberately chose ans\"atze of solutions and induced world volume metrics that did
reproduce those of the Kuperstein--Sonnenschein model, when we turned off the angular
velocity and world volume gauge fields. Moreover, in all examples, we demonstrated
that, independent from the value of angular velocity and world volume gauge fields,
in the large radii limit, the radial derivatives of the world volume field in our
ans\"atze coincide with that of vanishing angular velocity and vanishing world volume
gauge fields. Thus, in  all examples, in the large radii limit, our ans\"atze solved
to the brane field with asymptotic UV boundary values of the Kuperstein--Sonnenschein
model.  Furthermore, in all our examples, we demonstrated that, in the small radii
limit, in the IR, for specific range of angular velocities, the radial derivatives of
the world volume field in our ans\"atze behave as that of the Kuperstein--Sonnenschein
model, consistent with U-like embeddings.

The analysis of this paper may be extended in several ways. One immediate
extension is to study non-equilibrium steady states and energy dissipation
of flavored quarks that reside in fewer dimensions of spacetime. This involves
the computation of the induced world volume metrics and temperatures together
with the energy--stress tensors of lower dimensional rotating probe flavor
branes in \emph{U}-like embeddings in the Type IIB and/or Type IIA theory based
models. The other more demanding extension involves such analyses in confining
SUGRA backgrounds including IR deformations and additional background fluxes
which complicate the analytic form of probe brane solutions. We will consider
these extensions in subsequent future works.

\section*{Acknowledgement}

I am grateful to A.E.\,Mosaffa for encouragement and for collaboration
on previous related work. Thanks also to S.\,F.\,Taghavi for discussions.

\end{document}